# The competition between alpha decay and spontaneous fission in odd-even and odd-odd nuclei in the range $99 \leq Z \leq 129$

K. P. Santhosh* and B. Priyanka

*School of Pure and Applied Physics, Kannur University, Swami Anandatheertha Campus, Payyanur 670327, Kerala, INDIA*

email: drkpsanthosh@gmail.com

**Abstract.**

The predictions on the mode of decay of the odd-even and odd-odd isotopes of heavy and superheavy nuclei with $Z = 99\text{-}129$, in the range $228 \leq A \leq 336$, have been done within the Coulomb and proximity potential model for deformed nuclei (CPPMDN). A comparison of our calculated alpha half lives with the values computed using other theoretical models shows good agreement with each other. An extensive study on the spontaneous fission half lives of all the isotopes under study has been performed to identify the long-lived isotopes in the mass region. The study reveals that the alpha decay half lives and the mode of decay of the isotopes with $Z = 109, 111, 113, 115$ and $117$, evaluated using our formalisms, agrees well with the experimental observations. As our study on the odd-even and odd-odd isotopes of $Z = 99\text{-}129$ predicts that, the isotopes $^{238,240\text{-}254}99$, $^{244,246\text{-}258}101$, $^{248,250,252\text{-}260,262}103$, $^{254,256,258\text{-}262,264}105$, $^{258,260,262\text{-}264,266}107$, $^{262,264,266\text{-}274}109$, $^{266,268\text{-}279}111$, $^{270\text{-}284,286}113$, $^{272\text{-}289,291}115$, $^{274\text{-}299}117$, $^{276\text{-}307}119$, $^{281\text{-}314}121$, $^{287\text{-}320,322}123$, $^{295\text{-}325}125$, $^{302\text{-}327}127$ and $^{309\text{-}329}129$ survive fission and have alpha decay channel as the prominent mode of decay, these nuclei could possibly be synthesized in the laboratory and this could be of great interest to the experimentalists. The behaviour of these nuclei against the proton decay has also been studied to identify the probable proton emitters in this region of nuclei.

## 1. Introduction

Over the last thirty years, the experimentalists have launched an expedition to explore the predicted "*island of superheavy elements*", a region of increasingly stable nuclei around $Z = 114$, and so far, superheavy nuclei (SHN) with $Z \leq 118$ have been synthesised [1-18]. Though the availabilities and the advancements in stable nuclear beam technology have resulted in the fast growth of the nuclear chart especially in the superheavy (SH) mass region, except for an attempt [19] to produce $Z = 120$ superheavy nuclei through the $^{244}$Pu+$^{58}$Fe reaction, until now, any evidence on the production of nuclei with $Z > 118$ has not be obtained. While reviewing the production of new radioactive elements in the periodic table, it could be seen that, the "hot fusion" reactions or the "actinide-based fusion" reactions [2] performed mainly at JINR-FLNR, Dubna and the "cold fusion" reactions or the "cluster-based fusion" reactions [1, 3] performed mainly at GSI, Darmstadt, and at RIKEN, Japan, have shown impressive new prospects and these heavy-ion reactions have added significant contributions to our knowledge and understanding of nuclear properties. However, the short lifetimes and the low production cross sections observed in the fusion evaporation reactions increases the difficulty in the synthesis of new SHN and these poses difficulties to both experimentalists and theoreticians in studying the various properties of SH elements.

Even though alpha decay and spontaneous fission are the main decay modes of both heavy and SHN with $Z \geq 90$, spontaneous fission acts as the limiting factor that determines the stability of superheavy nuclei. In 1939, Bohr and Wheeler [20] predicted the phenomenon of spontaneous fission on the basis of liquid drop model and in 1940, Flerov et al [21] observed this phenomenon from $^{238}$U nucleus. Several empirical formulas have been put forward by several authors for determining the half lives of spontaneous fission and it was Swiatecki [22] in 1955, who proposed the first semi-empirical formula for spontaneous fission. At present, the spontaneous fission half lives of several superheavy nuclei have been measured in different laboratories [9, 10, 23-26] and extensive theoretical studies on the spontaneous fission half lives of SHN have also been performed, by various theoretical groups, for identifying the long-lived SH elements. Within the macroscopic-microscopic approach, the spontaneous fission properties of deformed even-even, odd-A and odd-odd superheavy nuclei with $Z = 104$-$120$, were analyzed by Smolanczuk et al., [27]. Later Muntain et al [28] computed the height of spontaneous fission barriers of $Z = 96$-$120$, within the macroscopic and microscopic model. A phenomenological formula for the spontaneous fission half lives of even-even nuclei in their ground state was generalized to both the case of odd nuclei and fission isomers in 2005 by Ren et al., [29, 30] and recently Xu et al [31] proposed an empirical formula for determining the spontaneous fission half lives of even-even nuclei.

While the process of spontaneous fission has a purely quantum tunnelling effect, the alpha decay is considered as a process where an alpha cluster penetrates the Coulomb barrier after its formation in the parent nucleus. Since the discovery of alpha decay by Rutherford [32, 33] in 1909, the mechanism of alpha decay has been a topic of constant study for both the experimentalists and theoreticians. It was in 1928 that George Gamow interpreted the phenomenon of alpha decay as a consequence of the quantum penetration of α particle [34] through the potential barrier (the "tunnelling effect") and later, E. U. Condon in collaboration with R. W. Gurney independently explained this phenomenon by means of wave mechanics [35]. The half lives of different radioactive decay such as alpha decay and spontaneous fission are the experimental signatures of the formation of SHN in fusion reaction. Hence, the calculations of these half lives are important in identifying the decay chains of SHN.

Even though the investigations on the alpha decay chains of the SHN can provide plenary information on the degree of stability of the SHN and their existence in nature, the studies on the competition of spontaneous fission and alpha decay on SHN is rather more important because those superheavy nuclei with relatively small alpha decay half lives compared to spontaneous fission half lives will survive fission and thus can be detected in the laboratory through alpha decay. Recently, Santhosh et al., have studied [36] the alpha decay, cluster decay and spontaneous fission in $^{294-326}$122 superheavy isotopes and the authors have also explored the competition between alpha decay and spontaneous fission in even-even superheavy isotopes with $Z = 100$-$122$, through the evaluation of alpha decay and spontaneous fission half lives of these isotopes [37]. In the present manuscript, as an extension of our previous studies on the competition between alpha decay and spontaneous fission of SHN [36, 37] and alpha decay chains of SHN [38-43], we have aimed at exploring the possibility of finding long-lived SH elements by comparing the alpha decay half lives, computed within the Coulomb and proximity potential model for deformed nuclei (CPPMDN) proposed by Santhosh et al., [44], with the spontaneous fission half lives of odd-even and odd-odd nuclei with $Z = 99$-$129$, so that the mode of decay of these nuclei could be predicted.

The details on the theoretical model used for the present study, CPPMDN, are presented in section 2. A detailed discussion on the alpha decay and spontaneous fission of the nuclei under study, the results obtained and a review on the various theoretical models used are given in section 3, and the section 4 presents a conclusion on the entire work.

## 2. The Coulomb and Proximity Potential Model for Deformed Nuclei (CPPMDN)

The interacting potential between two nuclei in CPPMDN is taken as the sum of deformed Coulomb potential, deformed two term proximity potential and centrifugal potential, for both the touching configuration and for the separated fragments. For the pre-scission (overlap) region, simple power law interpolation as done by Shi and Swiatecki [45] has been used.

The interacting potential barrier for two spherical nuclei is given by

$$V = \frac{Z_1 Z_2 e^2}{r} + V_p(z) + \frac{\hbar^2 \ell(\ell+1)}{2\mu r^2}, \quad \text{for } z > 0 \tag{1}$$

Here $Z_1$ and $Z_2$ are the atomic numbers of the daughter and emitted cluster, '$r$' is the distance between fragment centres, '$z$' is the distance between the near surfaces of the fragments, $\ell$ represents the angular momentum and $\mu$ the reduced mass. $V_P$ is the proximity potential given by Blocki et al., [46, 47] as,

$$V_p(z) = 4\pi \gamma b \left[ \frac{C_1 C_2}{(C_1 + C_2)} \right] \Phi\left(\frac{z}{b}\right) \tag{2}$$

with the nuclear surface tension coefficient,

$$\gamma = 0.9517 [1 - 1.7826 (N - Z)^2 / A^2] \text{ MeV/fm}^2 \tag{3}$$

Here $N$, $Z$ and $A$ represent the neutron, proton and mass number of the parent and $\Phi$ represents the universal proximity potential [47] given as

$$\Phi(\varepsilon) = -4.41 e^{-\varepsilon/0.7176}, \text{ for } \varepsilon \geq 1.9475 \tag{4}$$

$$\Phi(\varepsilon) = -1.7817 + 0.9270\varepsilon + 0.0169\varepsilon^2 - 0.05148\varepsilon^3, \text{ for } 0 \leq \varepsilon \leq 1.9475 \tag{5}$$

with $\varepsilon = z/b$, where the width (diffuseness) of the nuclear surface $b \approx 1$ fermi and the Süsmann central radii $C_i$ of the fragments are related to the sharp radii $R_i$ as

$$C_i = R_i - \left(\frac{b^2}{R_i}\right) \tag{6}$$

For $R_i$, we use semi-empirical formula in terms of mass number $A_i$ as [46]

$$R_i = 1.28 A_i^{1/3} - 0.76 + 0.8 A_i^{-1/3} \tag{7}$$

The potential for the internal part (overlap region) of the barrier is given as,

$$V = a_0(L-L_0)^n, \text{ for } z < 0 \qquad (8)$$

where $L = z + 2C_1 + 2C_2$ and $L_0 = 2C$, the diameter of the parent nuclei. The constants $a_0$ and $n$ are determined by the smooth matching of the two potentials at the touching point.

Using the one dimensional Wentzel-Kramers-Brillouin (WKB) approximation, the barrier penetrability $P$ is given as

$$P = \exp\left\{-\frac{2}{\hbar}\int_a^b \sqrt{2\mu(V-Q)}\,dz\right\} \qquad (9)$$

Here the mass parameter is replaced by $\mu = mA_1A_2/A$, where $m$ is the nucleon mass and $A_1$, $A_2$ are the mass numbers of daughter and emitted cluster respectively. The turning points "$a$" and "$b$" are determined from the equation, $V(a) = V(b) = Q$. The above integral can be evaluated numerically or analytically, and the half life time is given by

$$T_{1/2} = \left(\frac{\ln 2}{\lambda}\right) = \left(\frac{\ln 2}{\nu P}\right) \qquad (10)$$

where, $\nu = \left(\frac{\omega}{2\pi}\right) = \left(\frac{2E_v}{h}\right)$ represent the number of assaults on the barrier per second and $\lambda$ the decay constant. $E_v$, the empirical vibration energy is given as [48]

$$E_v = Q\left\{0.056 + 0.039\exp\left[\frac{(4-A_2)}{2.5}\right]\right\}, \quad \text{for } A_2 \geq 4 \qquad (11)$$

The Coulomb interaction between the two deformed and oriented nuclei with higher multipole deformation included [49, 50] is taken from Ref. [51] and is given as,

$$V_C = \frac{Z_1Z_2e^2}{r} + 3Z_1Z_2e^2\sum_{\lambda,i=1,2}\frac{1}{2\lambda+1}\frac{R_{0i}^\lambda}{r^{\lambda+1}}Y_\lambda^{(0)}(\alpha_i)\left[\beta_{\lambda i} + \frac{4}{7}\beta_{\lambda i}^2 Y_\lambda^{(0)}(\alpha_i)\delta_{\lambda,2}\right] \qquad (12)$$

with

$$R_i(\alpha_i) = R_{0i}\left[1 + \sum_\lambda \beta_{\lambda i} Y_\lambda^{(0)}(\alpha_i)\right] \qquad (13)$$

where $R_{0i} = 1.28A_i^{1/3} - 0.76 + 0.8A_i^{-1/3}$. Here $\alpha_i$ is the angle between the radius vector and symmetry axis of the $i^{\text{th}}$ nuclei (see Fig.1 of Ref [49]) and it is to be noted that the quadrupole interaction term proportional to $\beta_{21}\beta_{22}$, is neglected because of its short-range character.

The two-term proximity potential for interaction between a deformed and spherical nucleus is given by Baltz et al., [52] as

$$V_{P2}(R,\theta) = 2\pi \left[\frac{R_1(\alpha)R_C}{R_1(\alpha)+R_C+S}\right]^{1/2} \left[\frac{R_2(\alpha)R_C}{R_2(\alpha)+R_C+S}\right]^{1/2}$$

$$\times \left[\left[\varepsilon_0(S) + \frac{R_1(\alpha)+R_C}{2R_1(\alpha)R_C}\varepsilon_1(S)\right]\left[\varepsilon_0(S) + \frac{R_2(\alpha)+R_C}{2R_2(\alpha)R_C}\varepsilon_1(S)\right]\right]^{1/2} \quad (14)$$

where $R_1(\alpha)$ and $R_2(\alpha)$ are the principal radii of curvature of the daughter nuclei at the point where polar angle is $\alpha$, $R_C$ is the radius of the spherical cluster, $S$ is the distance between the surfaces along the straight line connecting the fragments and $\varepsilon_0(S)$ and $\varepsilon_1(S)$ are the one dimensional slab-on-slab function.

### 3. Results and Discussions

Authentic information on the degree of stability of the superheavy nuclei (SHN) and their existence in nature can be obtained through the investigations on the alpha decay chains of the SHN. But, the studies on the spontaneous fission and competition between spontaneous fission and alpha decay of SHN [29, 30, 38, 39, 53] has to be performed more lively because those superheavy nuclei with relatively small alpha decay half lives compared to the spontaneous fission half lives survive fission and thus can be detected in the laboratory via alpha decay. The present manuscript deals with the studies on the alpha decay properties of odd-even and odd-odd nuclei with $Z$ = 99-129, which has been done with an aim of exploring the possibility of finding the long-lived superheavy elements by comparing the calculated alpha decay half lives with the corresponding spontaneous fission half lives. The alpha decay studies have been performed within the Coulomb and proximity potential model for deformed nuclei (CPPMDN), where the external drifting potential is obtained as the sum of the deformed Coulomb potential, deformed two term proximity potential and centrifugal potential for the touching configuration and for the separated fragments. The reliability and applicability of CPPMDN in explaining the cluster decay and alpha decay have already been proved through our previous studies [54-58]. The energy released during the alpha transitions between the ground state energy levels of the parent nuclei and the ground state energy levels of the daughter nuclei, is given as,

$$Q_{gs \to gs} = \Delta M_p - (\Delta M_\alpha + \Delta M_d) + k(Z_p^\varepsilon - Z_d^\varepsilon) \quad (15)$$

Here, the terms $\Delta M_p$, $\Delta M_d$ and $\Delta M_\alpha$ represents the mass excess of the parent, daughter and the alpha particle respectively. The $Q$ values for the alpha decay have been evaluated using two different mass tables. For most of the nuclei under study, the recent experimental mass table of Wang *et al.*, [59] has been used and for those nuclei whose experimental mass excess were unavailable, the mass table of Koura-Tachibana-Uno-Yamada (KTUY) [60] has been used. As the effect of the atomic electrons on the energy of the alpha particle has not been included in the mass excess given in Ref. [59, 60], for a more accurate calculation of $Q$ value, the electron screening effect [61] has been included in equation (15). The term $k(Z_p^\varepsilon - Z_d^\varepsilon)$ represents this correction, where the terms $k$ = 8.7eV and $\varepsilon$ = 2.517 for nuclei with Z ≥ 60 and $k$ = 13.6eV and $\varepsilon$ = 2.408 for nuclei with Z < 60 have been derived from the data reported by Huang et al. [62]. The quadrupole ($\beta_2$) and hexadecapole ($\beta_4$) deformation values of both the parent and daughter nuclei have also been used for the evaluation of alpha half lives, and as the experimental deformation values were not available for the nuclei considered, the theoretical values taken from Ref. [63] have been used in the present manuscript.

### 3.1 Empirical Relations for Alpha Decay Half lives

For a theoretical comparison of our predicted alpha decay half lives with other theoretical models, the alpha decay half lives of all the isotopes under study have also been evaluated using the Viola-Seaborg semi-empirical relationship (VSS) [64], the analytical formulae of Royer [65] and the Universal (UNIV) curve of Poenaru et al., [66, 67].

The Viola-Seaborg semi-empirical relationship (VSS), with constants determined by Sobiczewski, Patyk and Cwiok [68], is given as,

$$\log_{10}(T_{1/2}) = (aZ+b)Q^{-1/2} + cZ + d + h_{\log} \qquad (16)$$

Here the half lives obtained are in seconds, Z is the atomic number of the parent nucleus and $Q$ represents the energy released during the reaction in MeV. The constants $a$ = 1.66175, $b$ = -8.5166, $c$ = -0.20228, $d$ = -33.9069 are adjustable parameters and the term $h_{\log}$ represents the hindrances associated with odd proton and odd neutron numbers, given by Viola-Seaborg [64].

During the latter half of the twentieth century, Royer studied the extent to which the generalized liquid drop model (GLDM) [69-71] could reproduce the α decay barrier and the alpha half lives, using the experimental $Q_\alpha$ value. The new, simple analytical formulae [65] for alpha decay of odd-even nuclei, is given as,

$$\log_{10}[T_{1/2}(s)] = -25.68 - 1.1423 A^{1/6}\sqrt{Z} + \frac{1.592Z}{\sqrt{Q_\alpha}} \tag{17}$$

and the relation for odd-odd nuclei is given as,

$$\log_{10}[T_{1/2}(s)] = -29.48 - 1.113 A^{1/6}\sqrt{Z} + \frac{1.6971Z}{\sqrt{Q_\alpha}} \tag{18}$$

To explain the decay half lives, Poenaru et al., derived the universal curves (UNIV) [66, 67, 72-74] by extending the fission theory to a larger mass asymmetry. Based on the quantum mechanical tunnelling process [75, 76], and using the decimal logarithm, the authors obtained the relation between the disintegration constant $\lambda$ and the partial decay half life $T$ of the parent nucleus as,

$$\lambda = \ln 2/T = \nu S P_s \tag{19}$$

$$\log_{10} T(s) = -\log_{10} P_s - \log_{10} S + [\log_{10}(\ln 2) - \log_{10} \nu] \tag{20}$$

The terms $\nu$, $S$ and $P_s$ are three model-dependent quantities, where $\nu$ is the frequency of assaults on the barrier per second, $S$ is the pre-formation probability of the cluster at the nuclear surface (equal to the penetrability of the internal part of the barrier in a fission theory [72, 73]), and $P_s$ is the quantum penetrability of the external potential barrier and the details on UNIV could be obtained from the Refs. [66, 67].

### 3.2 Empirical Relations for Spontaneous Fission Half lives

To identify the long-lived superheavy elements and the mode of decay of the isotopes under study, the spontaneous fission (SF) half lives of the corresponding nuclei have also been performed in the present work. The SF half lives have been evaluated using the semi-empirical relation of Santhosh et al., [77], the semi-empirical relation of Xu et al., [31] and the phenomenological formula of Ren et al., [29].

The new semi empirical formula for spontaneous fission developed by Santhosh et al., [77] is given as,

$$\log_{10}(T_{1/2}/yr) = a\frac{Z^2}{A} + b\left(\frac{Z^2}{A}\right)^2 + c\left(\frac{N-Z}{N+Z}\right) + d\left(\frac{N-Z}{N+Z}\right)^2 + e. \tag{21}$$

The terms $A$, $Z$ and $N$ in eqn. (21) represents respectively the mass number, atomic number and neutron number of the parent nuclei and the constants are $a$ = -43.25203, $b$ = 0.49192, $c$ = 3674.3927, $d$ = -9360.6 and $e$ = 580.75058.

For evaluating the spontaneous fission half lives of nuclei, Xu *et al.*, [31] formulated a simple semi empirical relation for even-even nuclei, and is given as,

$$T_{1/2} = \exp\left\{2\pi\left[C_0 + C_1 A + C_2 Z^2 + C_3 Z^4 + C_4(N-Z)^2 - (0.13323\frac{Z^2}{A^{1/3}} - 11.64)\right]\right\} \quad (22)$$

Here the half lives are obtained in seconds, $A$, $Z$ and $N$ represents respectively the mass number, atomic number and neutron number of the parent nuclei. The constants $C_0$ = -195.09227, $C_1$ = 3.10156, $C_2$ = -0.04386, $C_3$ = 1.4030x10$^{-6}$ and $C_4$ = -0.03199.

Ren et al., [29, 30] generalized a formula for spontaneous fission half-lives of even-even nuclei in their ground state to both cases of odd nuclei and of fission isomers [29] and the phenomenological formula proposed by Ren and Xu [29] is given by

$$\log_{10}(T_{1/2}/yr) = 21.08 + C_1\frac{(Z-90-v)}{A} + C_2\frac{(Z-90-v)^2}{A} + C_3\frac{(Z-90-v)^3}{A}$$

$$+ C_4\frac{(Z-90-v)(N-Z-52)^2}{A} \quad (23)$$

where $C_1$ = −548.825021, $C_2$ = −5.359139, $C_3$ = 0.767379 and $C_4$ = −4.282220. The seniority term $v$ was introduced taking the blocking effect of unpaired nucleon on the transfer of many nucleon-pairs during the fission process and $v = 0$ for spontaneous fission of even-even nuclei, $v = 2$ for odd $A$ and odd-odd nuclei.

In the present study, as mentioned before, only the odd $Z$, (both odd-even and odd-odd) nuclei have been considered. But, as the eqns. (21) and (22) were originally made to fit the even-even nuclei, instead of taking spontaneous fission half life, $T_{sf}$, directly, we have taken the average of fission half life, $T_{sf}^{av}$, of the corresponding neighboring even-even nuclei as the case may be. The $T_{sf}^{av}$ of two neighboring even-even nuclei and the $T_{sf}^{av}$ of four neighboring even-even nuclei have been considered while dealing with the odd-even nuclei and odd-odd nuclei respectively, as we have already performed in our earlier works [38]. Here we would like to mention that, in the case of $^{257}$99, $T_{sf}^{expt.}$ = 7.484×10$^5$s and $T_{sf}^{av}$ = 1.988×10$^5$s, in the case of $^{256}$101, $T_{sf}^{expt.}$ =1.64×10$^5$s and $T_{sf}^{av}$ = 9.574×10$^5$s and in the case of the nuclei of $^{259}$101, $T_{sf}^{expt.}$ = 5.760×10$^3$s and $T_{sf}^{av}$ = 9.044×10$^3$s, in the case of the nuclei $^{257}$103, $T_{sf}^{expt.}$ = 1.98×10$^3$s and $T_{sf}^{av}$ = 1.667×10$^3$s, in the case of $^{257}$105, $T_{sf}^{expt.}$ =

0.0348s and $T_{sf}^{av}$ = 0.0514s and in the case of $^{262}$105, $T_{sf}^{expt.}$ = 11.550s and $T_{sf}^{av}$ = 6.9415 s, which shows the good agreement between experimental [78] and computed average spontaneous fission half lives.

For the study on the possibilities of alpha decay of the odd-even and odd-odd heavy and superheavy nuclei with Z = 99-129, the isotopes $^{228-274}$99, $^{232-278}$101, $^{236-282}$103, $^{240-286}$105, $^{244-290}$107, $^{248-294}$109, $^{252-298}$111, $^{256-302}$113, $^{260-306}$115, $^{264-310}$117, $^{268-314}$119, $^{272-318}$121, $^{276-324}$123, $^{280-328}$125, $^{284-332}$127 and $^{288-336}$129 have been considered. The alpha decay half lives have also been calculated using the Viola-Seaborg semi-empirical relationship (VSS), the analytical formulae of Royer and the Universal curve (UNIV) of Poenaru et al., the formalisms described in section 3.1. Apart from this, we have also evaluated the alpha decay half lives within the Coulomb and proximity potential model (CPPM) formalism [79] (without the ground state deformation values of the both parent and daughter nuclei). For the comparative study of alpha decay versus spontaneous fission, the spontaneous fission half lives have been calculated using the semi-empirical relation of Santhosh et al., the semi-empirical relation of Xu et al., and the phenomenological formula of Ren et al., the formalisms discussed in section 3.2.

The entire study on the decay properties of the odd-even isotopes are given in the figures 1-4. Fig. 1 gives the plot for the $\log_{10}(T_{1/2})$ versus the mass number of the parent nuclei, of $^{229-273}$99, $^{233-277}$101, $^{237-281}$103 and $^{241-285}$105. From the comparison of the alpha decay half lives with the corresponding spontaneous fission half lives, it can be seen from fig. 3.1 (a), (b), (c) and (d) respectively that, the isotopes $^{241-253}$99, $^{247-257}$101, $^{253-259}$103 and $^{259,261}$105 are situated within the inverted parabola and hence they have alpha decay as the dominant decay channel. The plot for the $\log_{10}(T_{1/2})$ versus mass number of the parent nuclei, of $^{245-289}$107, $^{249-293}$109, $^{253-297}$111 and $^{257-301}$113 is given in fig. 2. As the isotopes $^{263}$107, $^{267-273}$109, $^{269-279}$111 and $^{271-283}$113 are seen within the inverted parabola, these isotopes survive fission and hence could be identified through alpha decay experiments. Figures 3 and 4 gives the $\log_{10}(T_{1/2})$ versus mass number of the parent nuclei, of $^{261-305}$115, $^{265-309}$117, $^{269-313}$119, $^{273-317}$121 and $^{277-323}$123, $^{281-327}$125, $^{285-331}$127, $^{289-333}$129 respectively. From fig. 3 (a), (b), (c) and (d), the isotopes $^{273-291}$115, $^{275-299}$117, $^{277-307}$119 and $^{281-313}$121 could be attributed to be surviving fission as they can be seen within the inverted parabola. In a similar manner, the isotopes $^{287-319}$123, $^{295-325}$125, $^{303-327}$127 and $^{309-329}$129 found within the parabola, in fig. 4 (a), (b), (c) and (d), have alpha decay channel as their dominant mode of decay.

The alpha decay half lives evaluated within the formalisms discussed in section 3.1 have also been depicted in these figures. It is noteworthy that the computed alpha half lives (within both CPPMDN and CPPM) matches well with the values evaluated using all these empirical formulas for alpha decay. From the figures it can also be seen that as the deformation values are included, the alpha decay half lives decreases.

It is to be noted that, some of these isotopes (for eg. $^{275}$109, $^{279,281}$111, $^{283,285}$113, $^{287,289}$115 and $^{293}$117) in the superheavy region have already been synthesized experimentally and have been identified via alpha decay. So, in table 1, we have given a comparison of the computed alpha decay half lives with the experimental alpha half lives [14, 18], for the isotopes that have been experimentally detected. It can be seen that both the experimental alpha half lives and the alpha half lives computed within both CPPM and CPPMDN are in good agreement with each other, which proves the reliability of both CPPM and CPPMDN.

In tables 2-6, we have presented the predictions on the mode of decay of the isotopes of odd-even superheavy nuclei with $Z$ = 109, 111, 113, 115 and 117, those nuclei that have already been synthesized experimentally. In these tables those isotopes whose mass excess values were unavailable from the mass tables [59, 60] have not been included. The considered isotope and the $Q$ values for the corresponding alpha decays are given in column 1 and 2 respectively. In column 3 and 4 we have given the average spontaneous fission half lives computed using the phenomenological formula of Xu et al., and the semi-empirical relation of Santhosh et al. The calculations of the alpha decay half lives done within the CPPMDN formalism (with the ground state deformation values of the both parent and daughter nuclei) are given in column 5 and those calculated using the analytical formula of Royer, are given in column 6. The half life values computed using the VSS systematic is given in column 7, and in column 8, the half life values computed using UNIV have been presented. The predictions on the mode of decay of the isotopes under study have been given in column 9.

The entire study on the decay properties of the odd-odd isotopes have been presented in figures 5-8. The plot for the $\log_{10}(T_{1/2})$ versus mass number of the parent nuclei, of $^{228-274}$99, $^{232-278}$101, $^{236-282}$103 and $^{240-286}$105 is given in Fig. 5. The comparison of the alpha decay half lives with the corresponding spontaneous fission half lives shows that, the isotopes $^{238-254}$99, $^{244-258}$101, $^{248-262}$103 and $^{254-264}$105 in fig. 5 (a), (b), (c) and (d) respectively, are situated within the inverted parabola and hence they have alpha decay as the dominant decay channel. In fig. 6, the plot for the

$\log_{10}(T_{1/2})$ versus mass number of the parent nuclei, of $^{244-290}107$, $^{248-294}109$, $^{252-298}111$ and $^{256-302}113$ is given. As the isotopes $^{258-266}107$, $^{262-274}109$, $^{266-278}111$ and $^{270-286}113$ are seen within the inverted parabola, these isotopes survive fission and hence could be identified through alpha decay experiments. Figures 7 and 8 gives the $\log_{10}(T_{1/2})$ versus mass number of the parent nuclei, of $^{260-306}115$, $^{264-310}117$, $^{268-314}119$, $^{272-318}121$ and $^{276-324}123$, $^{280-328}125$, $^{284-332}127$, $^{288-336}129$ respectively. The isotopes $^{272-288}115$, $^{274-298}117$, $^{276-306}119$ and $^{282-314}121$ in fig. 7 (a), (b), (c) and (d) and the isotopes $^{288-322}123$, $^{296-324}125$, $^{302-326}127$ and $^{310-328}129$ in fig. 8 (a), (b), (c) and (d) could be attributed as to survive fission as they can be seen within the inverted parabola and hence have alpha decay channel as their dominant mode of decay.

It is noteworthy that some of these isotopes (for eg. $^{270,276,278}109$, $^{280,282}111$, $^{284}113$, $^{288}115$ and $^{294}117$) in the superheavy region have already been synthesized experimentally and have been identified via alpha decay. Table 7 gives a comparison of the computed alpha decay half lives with the experimental alpha half lives [14, 18], for the isotopes that have been experimentally detected. The reliability of both CPPM and CPPMDN, could be revealed from the fact the both the experimental alpha half lives and the alpha half lives computed within both CPPM and CPPMDN are in good agreement with each other.

The predictions on the mode of decay of the isotopes of the odd-odd superheavy nuclei with Z = 109, 111, 113, 115 and 117, those nuclei that have already been synthesized experimentally, have been presented in the tables 8-12. The considered isotope and the $Q$ values for the corresponding alpha decays are given in column 1 and 2 respectively. In column 3 and 4 we have given the average spontaneous fission half lives computed using the phenomenological formula of Xu et al., and the semi-empirical relation of Santhosh et al. The calculations of the alpha decay half lives done within the CPPMDN formalism (with the ground state deformation values of the both parent and daughter nuclei) are given in column 5 and those calculated using the analytical formula of Royer, are given in column 6. The half life values computed using the VSS systematic is given in column 7 and in column 8, the half life values computed using UNIV have been presented. The predictions on the mode of decay of the isotopes under study have been given in column 9.

The tables 2-6 and the tables 8-12 predicts the alpha half lives and the mode of decay of some superheavy isotopes, with Z = 109, Z = 111, Z = 113, Z = 115 and Z = 117, most of which presented in the tables are for unknown isotopes. Hence, we hope that these predictions to help and guide the experimentalists for the synthesis of new superheavy isotopes.

The Seaborg plots connecting the spontaneous fission half lives and the fissionability parameter ($Z^2/A$) for the odd-even and the odd-odd isotopes with $Z = 99-129$ are given in the figures 9 and 10 respectively. It is to be noted that, generally, for a given value of $Z$, the spontaneous fission half lives increases with increasing fissionability parameter, reaches a maximum and then decreases with increasing fissionability parameter [37]. From the figures 9 and 10, the observation that for a given value of $Z$, the spontaneous fission half lives increases with increasing fissionability parameter, reaches a maximum and then decreases with increasing fissionability parameter, reveals the agreement with the general trend. Studier et al., [80] and Bonetti et al., [81] have studied the Seaborg plots for the isotopes with $Z = 90-100$ which shows that the spontaneous fission half lives exponentially decreases with increasing the proton number, $Z$. But, in our present study, from the figures 9 and 10, it can be seen that, as $Z$ increases, up to $Z = 109$, the spontaneous fission half lives decreases with the proton number $Z$, and then increases, which is a slight increase for $Z = 111-117$ and a sharp increase for $Z = 119-129$. For a fixed neutron number, the neutron separation energy increases with increasing proton number. The increase in the proton number $Z$ increases the fissility parameter which pulls in maximum of the liquid drop energy. So when protons are added beyond the shell closure $Z = 120$ (see [36]), the barrier height increases gradually which results in the reduced fission probability (increased half life). In general, the alpha disintegration half lives increases and the spontaneous fission half lives decreases beyond $Z = 120$. It is to be noted that, the trend observed in the present work deviates from the earlier observations [80].

The ratio of the Coulomb energy of a spherical sharp surface drop to twice the nuclear surface energy is defined as the fissility, $X = \dfrac{E_C^{(0)}}{2E_S^{(0)}}$, of a nucleus and it scales with $Z^2/A$, the fissionability parameter. The fissility parameter is normalized to critical fissility beyond which nuclei promptly disintegrate and is given as

$$X = \dfrac{Z^2/A}{50.883\left(1 - 1.7826\left(\dfrac{N-Z}{A}\right)^2\right)} \tag{24}$$

Figures 11 and 12 represents respectively the plot connecting the logarithmic ratio of spontaneous fission to alpha decay half lives against the fissility parameter for odd-even and odd-odd isotopes of $Z = 99-129$. The competition between spontaneous fission and alpha decay against the fissility parameter can be clearly seen from these plots. From these plots, we could see that for a given Z value, the ratio of spontaneous fission to alpha decay half lives increases with increasing fissility parameter, reaches a maximum and then decreases. It is also clear that the ratio of spontaneous fission to alpha decay half-lives increases with increasing proton number. But as Z increases, up to Z = 109, the logarithmic ratio of spontaneous fission half lives to the alpha half lives decreases with the Z, and then increases, which is a slight increase for Z = 111-117 and a sharp increase for Z = 119-129. The present observation indicates to the fact that, as the proton number increases, alpha decay becomes the most dominant mode of decay.

The behaviour of logarithmic ratio of spontaneous fission to alpha decay half lives against the relative neutron excess, $I = \frac{N-Z}{N+Z}$ for the odd-even and odd-odd isotopes of Z = 99-129 has also been studied in the present manuscript and is given as figures 13 and 14 respectively. The spontaneous fission half lives evaluated using the phenomenological formula of Ren at al., has been used for plotting these graphs. From these plots it can be seen that, the curves converge at 0.2 and then diverges. So to say, those isotopes with neutron excess less than or equal to 0.2 can be considered as the probable candidates that can survive spontaneous fission. Thus the point, which corresponds to the value Z/A = 0.4, can be treated as the limit for spontaneous fission. As it is evident from the figures that, the isotopes in the first quadrant survive fission and have alpha decay as the dominant mode of decay, these isotopes can thus be synthesized and identified via alpha decay in the laboratories.

As our theoretical studies on the competition of alpha decay and spontaneous fission of odd-even and odd-odd heavy and superheavy nuclei with Z = 99-129 reveals some interesting facts regarding those isotopes situated within the inverted parabola, we were also interested in studying how these isotopes behaved to proton decay. Thus, in order to identify the proton emitters among these nuclei, the one-proton and the two-proton separation energies [82] of all the isotopes under study were evaluated using the relations given as

$$S(p) = -\Delta M(A,Z) + \Delta M(A-1, Z-1) + \Delta M_H = -Q(\gamma, p) \quad (25)$$

$$S(2p) = -\Delta M(A,Z) + \Delta M(A-2, Z-2) + 2\Delta M_H = -Q(\gamma, 2p) \quad (26)$$

where, the terms $S(p)$ and $S(2p)$ represents the one-proton separation energy and two-proton separation energy of the nuclei respectively. The term $\Delta M(A,Z)$ represents the mass excess of the parent and $\Delta M_H$ represents the mass excess of the proton. The mass excess of the daughter nuclei produced during the one-proton and the two-proton radioactivities are represented by the terms $\Delta M(A-1, Z-1)$ and $\Delta M(A-2, Z-2)$ respectively. The terms $Q(\gamma, p)$ and $Q(\gamma, 2p)$ represents respectively the $Q$ values for the one-proton radioactivity and the two-proton radioactivity. Here we would like to mention that, all those isotopes with negative values of proton separation energies, lie outside the proton drip line and thus may easily decay through proton emission. On calculating the proton separation energies for the odd-even and odd-odd isotopes of $Z$ = 99-129, we could see some more interesting facts. Our calculations reveal that the isotopes, $^{262,264}$109, $^{266,268-271}$111, $^{270-277}$112, $^{272-281}$115, $^{274-287}$119, $^{276-293}$119, $^{281-301}$121, $^{287-309}$123, $^{295-315}$125, $^{302-321}$127 and $^{309-327}$129, with $Z$ = 109-129 have negative values of proton separation energies. Our studies on the alpha and spontaneous fission half lives have already revealed these isotopes to be situated within the inverted parabola. The study on the proton separation energies of these nuclei reveals that these isotopes may also decay through proton emission.

Thus, it is noteworthy that, the manuscript presents an extensive study on the competition of alpha decay and spontaneous fission of the odd-even and odd-odd isotopes of $Z$ = 99-129 and through a fruitful comparison of the evaluated alpha half lives with both the spontaneous fission and proton decay calculations, we could identify the range of nuclei that decays through these three decay modes. We have also predicted the mode of decay of a wide range of unknown isotopes of both the heavy and superheavy nuclei and we would also like to mention that, the presented work may provide a new ray of hope to the experimentalists, for the synthesis of new isotopes.

## 4. Conclusion

Within the Coulomb and proximity potential model for deformed nuclei (CPPMDN), the alpha decay half lives of 371 odd-even isotopes and 388 odd-odd isotopes of $Z$ = 99-129 within the mass range 228 ≤ A ≤ 336 have been evaluated. The alpha decay half lives have also been calculated using the Viola-Seaborg semi-empirical relationship (VSS), the analytical formula of Royer and the Universal formula of Poenaru et al. A comparison of the predicted half lives with the values evaluated using these empirical formulas are in agreement with each other and hence both CPPM and CPPMDN could also be considered as unified models for alpha and cluster decay studies. The spontaneous fission half lives have been evaluated using the semi-empirical formula of Santhosh et al., the semi-empirical relation given by Xu et al., and the phenomenological formula of Ren et al., so as to predict the mode of decay of the isotopes under study. A comparative study on the competition of alpha decay versus spontaneous fission on the odd-even and odd-odd isotopes of

heavy and superheavy nuclei reveals that the isotopes $^{238,\,240-254}99$, $^{244,\,246-258}101$, $^{248,\,250,\,252-260,262}103$, $^{254,\,256,\,258-262,\,264}105$, $^{258,\,260,\,262-264,\,266}107$, $^{262,\,264,\,266-274}109$, $^{266,\,268-279}111$, $^{270-284,\,286}113$, $^{272-289,\,291}115$, $^{274-299}117$, $^{276-307}119$, $^{281-314}121$, $^{287-320,\,322}123$, $^{295-325}125$, $^{302-327}127$ and $^{309-329}129$ are situated within the inverted parabola. Thus these isotopes survive fission and have alpha decay channel as the prominent mode of decay and are possible to be synthesized in the laboratory. As most of these isotopes have already been synthesized and identified via alpha decay in the laboratory, we hope our predictions to be more helpful to the experimentalists. The plots of the spontaneous fission half lives against the fissionability parameter, the plots on the ratio of spontaneous fission half lives to the alpha half lives against the fissility parameter and the behaviour of logarithmic ratio of spontaneous fission to alpha decay half lives against the relative neutron excess, of both odd-even and odd-odd isotopes, have been studied in detail and this indicates that, as the proton number increases, alpha decay becomes the most dominant mode of decay. The present study thus reveals the competition between spontaneous fission and alpha decay. The behaviour of all these isotopes against proton decay has also been studied and hence we could identify those isotopes which lie outside the proton drip line and thus decaying through proton emission. Thus, we presume that the present study and the theoretical predictions on the mode of decay of a wide range of odd-even and odd-odd isotopes of $Z = 99$-$129$ to provide a ray of hope to enhance future experiments.

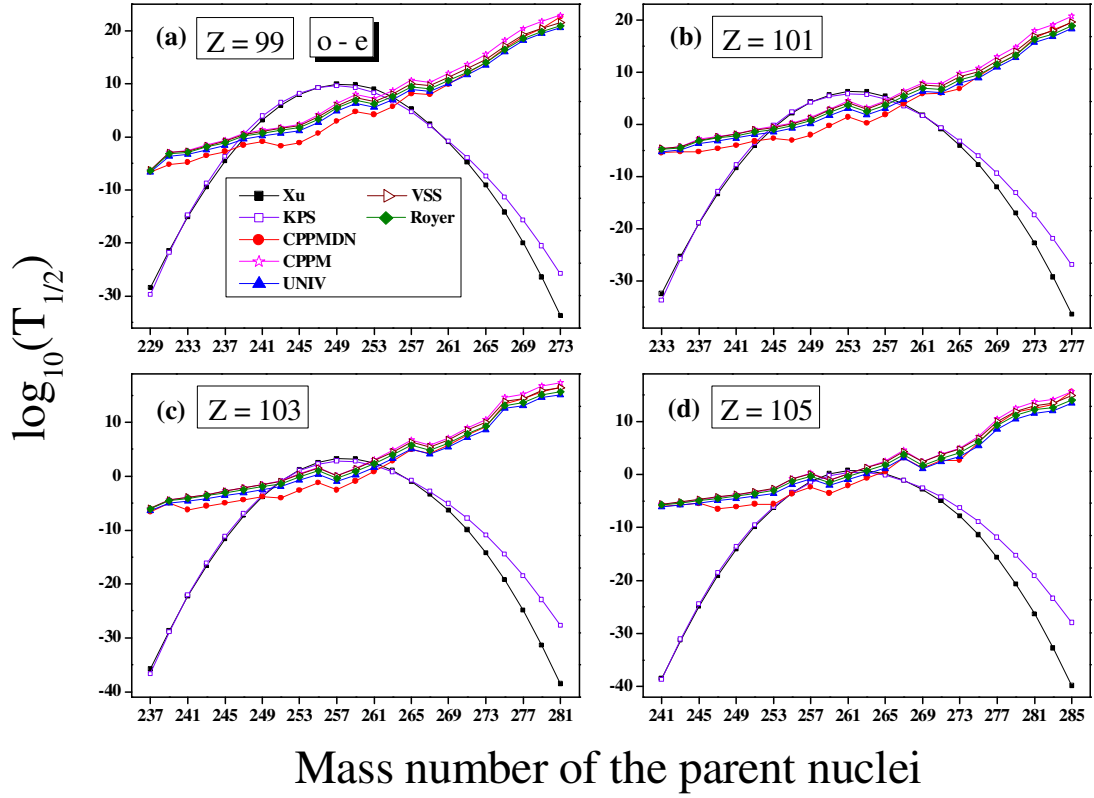

**Figure 1.** (Color online) Plot for the comparison of the computed alpha decay half lives with the spontaneous fission half lives for the odd-even, (a) $^{229-273}$99 (b) $^{233-277}$101 (c) $^{237-281}$103 and (d) $^{241-285}$105, isotopes.

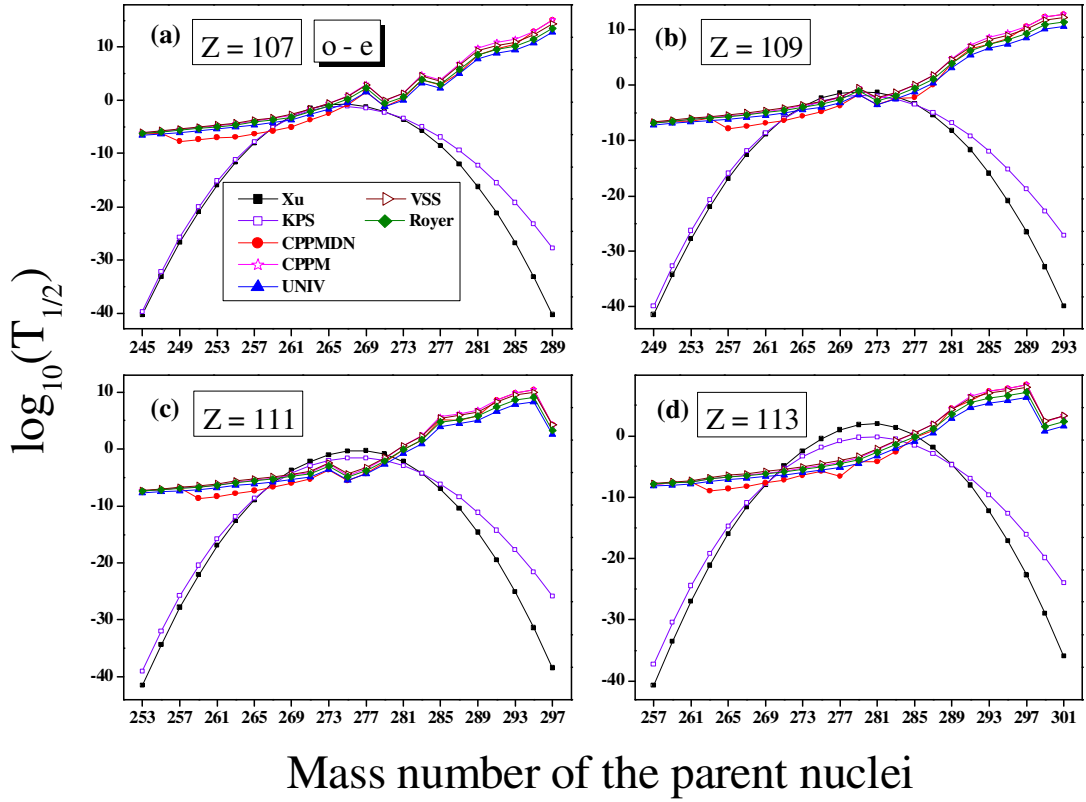

**Figure 2.** (Color online) Plot for the comparison of the computed alpha decay half lives with the spontaneous fission half lives for the odd-even, (a) $^{245-289}$107 (b) $^{249-293}$109 (c) $^{253-297}$111 and (d) $^{257-301}$113, isotopes.

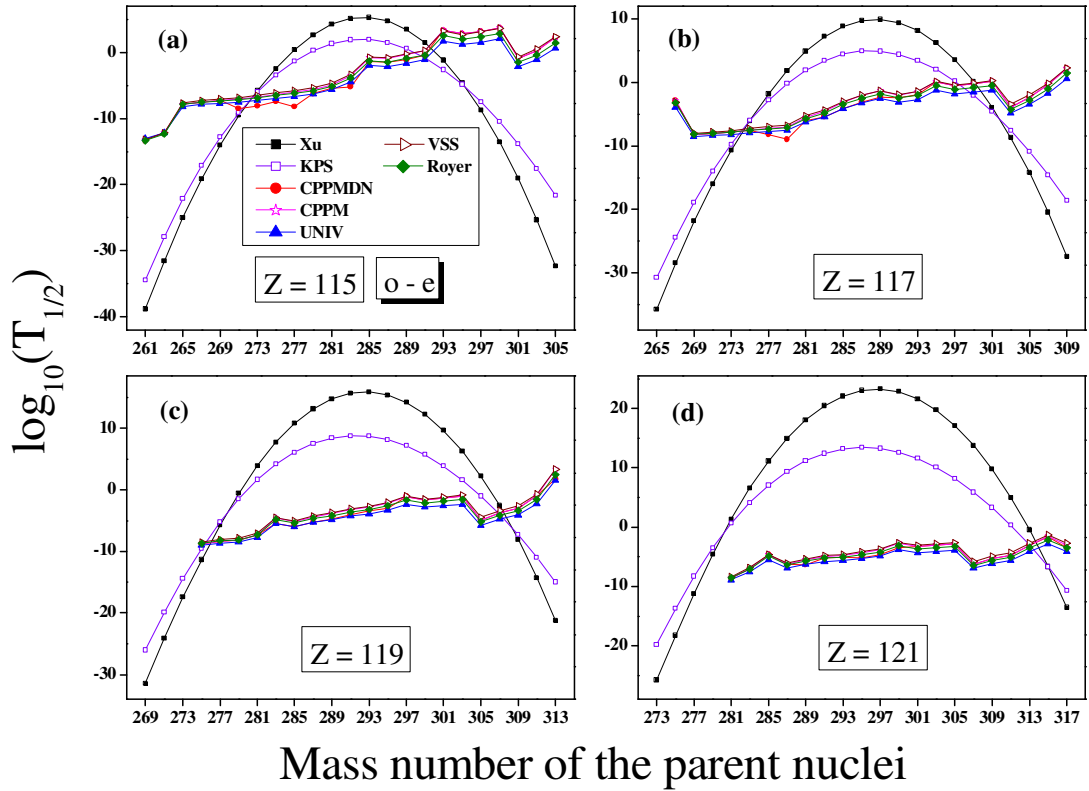

**Figure 3.** (Color online) Plot for the comparison of the computed alpha decay half lives with the spontaneous fission half lives for the odd-even, (a) $^{261-305}$115 (b) $^{265-309}$117 (c) $^{269-313}$119 and (d) $^{273-317}$121, isotopes.

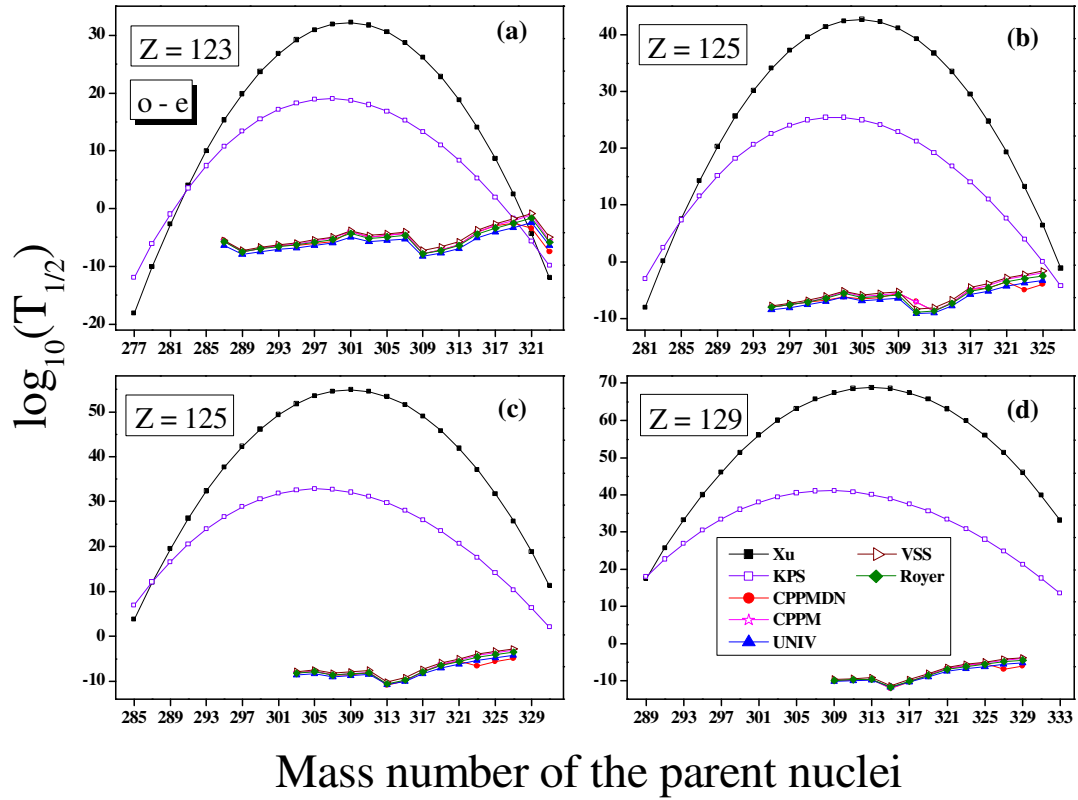

**Figure 4.** (Color online) Plot for the comparison of the computed alpha decay half lives with the spontaneous fission half lives for the odd-even (a) $^{277-323}$123 (b) $^{281-327}$125 (c) $^{285-331}$127 and (d) $^{289-333}$129 isotopes.

**Table 1.** A comparison of the alpha decay half lives of the experimentally synthesised odd-even superheavy nuclei with the corresponding experimental alpha half lives [14, 18]. The alpha half lives calculations are done for zero angular momentum transfers.

| Parent Nuclei | $Q_\alpha$ (cal.) MeV | $T_{1/2}^\alpha$ (s) (Expt.) | $T_{1/2}^\alpha$ (s) | | | | |
|---|---|---|---|---|---|---|---|
| | | | CPPMDN | CPPM | VSS | UNIV | Royer |
| $^{275}$109 | 10.268 | 9.700x10$^{-3}$ | 4.107x10$^{-3}$ | 3.420x10$^{-2}$ | 4.832x10$^{-2}$ | 2.910x10$^{-3}$ | 1.149x10$^{-2}$ |
| $^{279}$111 | 10.570 | 1.698x10$^{-1}$ | 5.386x10$^{-3}$ | 2.386x10$^{-2}$ | 3.372x10$^{-2}$ | 1.956x10$^{-3}$ | 8.131x10$^{-3}$ |
| $^{281}$111 | 9.820 | 2.600x10$^{1}$ | 7.648x10$^{-1}$ | 3.384x10$^{0}$ | 3.602x10$^{0}$ | 1.675x10$^{-1}$ | 8.147x10$^{-1}$ |
| $^{283}$113 | 10.541 | 1.000x10$^{-1}$ | 2.654x10$^{-3}$ | 1.383x10$^{-1}$ | 1.660x10$^{-1}$ | 8.709x10$^{-3}$ | 4.117x10$^{-2}$ |
| $^{285}$113 | 10.091 | 5.500x10$^{0}$ | 5.381x10$^{-1}$ | 2.617x10$^{0}$ | 2.740x10$^{0}$ | 1.221x10$^{-1}$ | 6.309x10$^{-1}$ |
| $^{287}$115 | 10.793 | 3.200x10$^{-2}$ | 3.462x10$^{-2}$ | 1.306x10$^{-1}$ | 1.514x10$^{-1}$ | 7.674x10$^{-3}$ | 3.860x10$^{-2}$ |
| $^{289}$115 | 10.573 | 2.200x10$^{-1}$ | 8.647x10$^{-2}$ | 5.030x10$^{-1}$ | 5.692x10$^{-1}$ | 2.557x10$^{-2}$ | 1.340x10$^{-1}$ |
| $^{293}$117 | 11.244 | 1.400x10$^{-2}$ | 1.147x10$^{-2}$ | 3.360x10$^{-2}$ | 4.357x10$^{-2}$ | 2.091x10$^{-3}$ | 1.056x10$^{-2}$ |

**Table 2.** The alpha decay half lives and the spontaneous fission half lives of $^{249-293}$109 isotopes. The mode of decay is predicted by comparing the alpha decay half lives with the spontaneous fission half lives. The alpha half lives calculations are done for zero angular momentum transfers.

| Parent nuclei | $Q_\alpha$ (cal.) MeV | $T_{SF}^{av}$ (Xu) (s) | $T_{SF}^{av}$ (KPS) (s) | $T_{1/2}^\alpha$ (s) | | | | Mode of Decay |
|---|---|---|---|---|---|---|---|---|
| | | | | CPPMDN | Royer | VSS | UNIV | |
| $^{249}$109 | 12.658 | 3.721x10$^{-42}$ | 1.314x10$^{-40}$ | 1.660x10$^{-7}$ | 1.514x10$^{-7}$ | 2.153x10$^{-7}$ | 6.512x10$^{-8}$ | SF |
| $^{251}$109 | 12.488 | 6.017x10$^{-35}$ | 2.334x10$^{-33}$ | 3.426x10$^{-7}$ | 2.959x10$^{-7}$ | 4.593x10$^{-7}$ | 1.222x10$^{-7}$ | SF |
| $^{253}$109 | 12.318 | 1.865x10$^{-28}$ | 5.639x10$^{-27}$ | 3.380x10$^{-7}$ | 5.878x10$^{-7}$ | 9.953x10$^{-7}$ | 2.331x10$^{-7}$ | SF |
| $^{255}$109 | 12.178 | 1.109x10$^{-22}$ | 2.022x10$^{-21}$ | 1.031x10$^{-6}$ | 1.031x10$^{-6}$ | 1.905x10$^{-6}$ | 3.964x10$^{-7}$ | SF |
| $^{257}$109 | 12.018 | 1.265x10$^{-17}$ | 1.171x10$^{-16}$ | 1.315x10$^{-8}$ | 2.014x10$^{-6}$ | 4.055x10$^{-6}$ | 7.466x10$^{-7}$ | SF |
| $^{259}$109 | 12.848 | 2.774x10$^{-13}$ | 1.185x10$^{-12}$ | 3.734x10$^{-8}$ | 4.197x10$^{-6}$ | 9.203x10$^{-6}$ | 1.496x10$^{-6}$ | SF |
| $^{261}$109 | 11.648 | 1.169x10$^{-9}$ | 2.264x10$^{-9}$ | 1.272x10$^{-7}$ | 1.036x10$^{-5}$ | 2.470x10$^{-5}$ | 3.522x10$^{-6}$ | SF |
| $^{263}$109 | 11.438 | 9.492x10$^{-7}$ | 8.793x10$^{-7}$ | 3.602x10$^{-7}$ | 2.763x10$^{-5}$ | 7.158x10$^{-5}$ | 8.945x10$^{-6}$ | SF |
| $^{265}$109 | 11.168 | 1.490x10$^{-4}$ | 7.470x10$^{-5}$ | 2.756x10$^{-6}$ | 1.047x10$^{-4}$ | 2.938x10$^{-4}$ | 3.176x10$^{-5}$ | SF |
| $^{267}$109 | 10.918 | 4.621x10$^{-3}$ | 1.506x10$^{-3}$ | 1.470x10$^{-5}$ | 3.741x10$^{-4}$ | 1.138x10$^{-3}$ | 1.072x10$^{-4}$ | α |
| $^{269}$109 | 10.578 | 3.125x10$^{-2}$ | 8.075x10$^{-3}$ | 1.921x10$^{-4}$ | 2.359x10$^{-3}$ | 7.743x10$^{-3}$ | 6.256x10$^{-4}$ | α |
| $^{271}$109 | 9.958 | 6.248x10$^{-2}$ | 1.396x10$^{-2}$ | 1.913x10$^{-2}$ | 9.354x10$^{-2}$ | 3.282x10$^{-1}$ | 2.168x10$^{-2}$ | α |
| $^{273}$109 | 10.658 | 4.632x10$^{-2}$ | 9.738x10$^{-3}$ | 2.646x10$^{-4}$ | 1.251x10$^{-3}$ | 4.891x10$^{-3}$ | 3.456x10$^{-4}$ | α |
| $^{275}$109 | 10.268 | 9.492x10$^{-3}$ | 2.831x10$^{-3}$ | 4.107x10$^{-3}$ | 1.149x10$^{-2}$ | 4.832x10$^{-2}$ | 2.910x10$^{-3}$ | SF |
| $^{277}$109 | 9.768 | 4.127x10$^{-4}$ | 3.019x10$^{-4}$ | 5.718x10$^{-3}$ | 2.468x10$^{-1}$ | 1.110x10$^{0}$ | 5.637x10$^{-2}$ | SF |
| $^{279}$109 | 9.188 | 3.533x10$^{-6}$ | 1.101x10$^{-5}$ | 1.181x10$^{0}$ | 1.206x10$^{1}$ | 5.780x10$^{1}$ | 2.486x10$^{0}$ | SF |
| $^{281}$109 | 8.348 | 5.869x10$^{-9}$ | 1.381x10$^{-7}$ | 8.676x10$^{3}$ | 7.179x10$^{3}$ | 3.616x10$^{4}$ | 1.326x10$^{3}$ | SF |
| $^{283}$109 | 7.738 | 1.887x10$^{-12}$ | 6.151x10$^{-10}$ | 3.752x10$^{6}$ | 1.387x10$^{6}$ | 7.383x10$^{6}$ | 2.461x10$^{5}$ | SF |
| $^{285}$109 | 7.418 | 1.175x10$^{-16}$ | 1.014x10$^{-12}$ | 2.542x10$^{7}$ | 2.739x10$^{7}$ | 1.558x10$^{8}$ | 4.839x10$^{6}$ | SF |
| $^{287}$109 | 7.238 | 1.418x10$^{-21}$ | 6.445x10$^{-16}$ | 2.884x10$^{8}$ | 1.546x10$^{8}$ | 9.454x10$^{8}$ | 2.744x10$^{7}$ | SF |
| $^{289}$109 | 6.988 | 3.314x10$^{-27}$ | 1.647x10$^{-19}$ | 3.859x10$^{10}$ | 1.983x10$^{9}$ | 1.298x10$^{10}$ | 3.558x10$^{8}$ | SF |
| $^{291}$109 | 6.658 | 1.503x10$^{-33}$ | 1.759x10$^{-23}$ | 1.850x10$^{12}$ | 7.396x10$^{10}$ | 5.148x10$^{11}$ | 1.361x10$^{10}$ | SF |
| $^{293}$109 | 6.558 | 1.322x10$^{-40}$ | 8.157x10$^{-28}$ | 6.064x10$^{12}$ | 2.211x10$^{11}$ | 1.659x10$^{12}$ | 4.126x10$^{10}$ | SF |

**Table 3.** The alpha decay half lives and the spontaneous fission half lives of $^{253-297}111$ isotopes. The mode of decay is predicted by comparing the alpha decay half lives with the spontaneous fission half lives. The alpha half lives calculations are done for zero angular momentum transfers.

| Parent nuclei | $Q_\alpha$ (cal. MeV) | $T^{av}_{SF}$ (Xu) (s) | $T^{av}_{SF}$ (KPS) (s) | $T^\alpha_{1/2}$ (s) | | | | Mode of Decay |
|---|---|---|---|---|---|---|---|---|
| | | | | CPPMDN | Royer | VSS | UNIV | |
| $^{253}111$ | 13.240 | 2.538x10$^{-42}$ | 8.870x10$^{-40}$ | 4.178x10$^{-8}$ | 4.156x10$^{-8}$ | 5.813x10$^{-8}$ | 1.874x10$^{-8}$ | SF |
| $^{255}111$ | 13.090 | 4.378x10$^{-35}$ | 9.652x10$^{-33}$ | 7.555x10$^{-8}$ | 7.185x10$^{-8}$ | 1.098x10$^{-7}$ | 3.126x10$^{-8}$ | SF |
| $^{257}111$ | 12.940 | 1.448x10$^{-28}$ | 1.530x10$^{-26}$ | 1.386x10$^{-7}$ | 1.257x10$^{-7}$ | 2.097x10$^{-7}$ | 5.277x10$^{-8}$ | SF |
| $^{259}111$ | 12.820 | 9.183x10$^{-23}$ | 3.841x10$^{-21}$ | 2.033x10$^{-9}$ | 1.947x10$^{-7}$ | 3.549x10$^{-7}$ | 7.972x10$^{-8}$ | SF |
| $^{261}111$ | 12.620 | 1.118x10$^{-17}$ | 1.654x10$^{-16}$ | 4.542x10$^{-9}$ | 4.365x10$^{-7}$ | 8.668x10$^{-7}$ | 1.698x10$^{-7}$ | SF |
| $^{263}111$ | 12.400 | 2.616x10$^{-13}$ | 1.318x10$^{-12}$ | 1.617x10$^{-8}$ | 1.098x10$^{-6}$ | 2.373x10$^{-6}$ | 4.036x10$^{-7}$ | SF |
| $^{265}111$ | 12.220 | 1.178x10$^{-9}$ | 2.097x10$^{-9}$ | 4.770x10$^{-8}$ | 2.346x10$^{-6}$ | 5.520x10$^{-6}$ | 8.260x10$^{-7}$ | SF |
| $^{267}111$ | 12.020 | 1.029x10$^{-6}$ | 7.191x10$^{-7}$ | 2.227x10$^{-7}$ | 5.635x10$^{-6}$ | 1.442x10$^{-5}$ | 1.890x10$^{-6}$ | SF |
| $^{269}111$ | 11.800 | 1.801x10$^{-4}$ | 5.824x10$^{-5}$ | 9.257x10$^{-7}$ | 1.534x10$^{-5}$ | 4.264x10$^{-5}$ | 4.876x10$^{-6}$ | α |
| $^{271}111$ | 11.570 | 7.346x10$^{-3}$ | 1.279x10$^{-3}$ | 5.318x10$^{-6}$ | 4.538x10$^{-5}$ | 1.369x10$^{-4}$ | 1.365x10$^{-5}$ | α |
| $^{273}111$ | 10.960 | 9.903x10$^{-2}$ | 9.365x10$^{-3}$ | 2.369x10$^{-4}$ | 1.111x10$^{-3}$ | 3.601x10$^{-3}$ | 2.860x10$^{-4}$ | α |
| $^{275}111$ | 11.790 | 4.715x10$^{-1}$ | 2.688x10$^{-2}$ | 2.829x10$^{-6}$ | 1.245x10$^{-5}$ | 4.483x10$^{-5}$ | 4.084x10$^{-6}$ | α |
| $^{277}111$ | 11.290 | 5.580x10$^{-1}$ | 2.902x10$^{-2}$ | 5.652x10$^{-5}$ | 1.532x10$^{-4}$ | 5.942x10$^{-4}$ | 4.401x10$^{-5}$ | α |
| $^{279}111$ | 10.570 | 1.351x10$^{-1}$ | 1.064x10$^{-2}$ | 5.386x10$^{-3}$ | 8.131x10$^{-3}$ | 3.372x10$^{-2}$ | 1.956x10$^{-3}$ | α |
| $^{281}111$ | 9.820 | 6.390x10$^{-3}$ | 1.285x10$^{-3}$ | 7.648x10$^{-1}$ | 8.147x10$^{-1}$ | 3.602x10$^{0}$ | 1.675x10$^{-1}$ | SF |
| $^{283}111$ | 9.240 | 5.852x10$^{-5}$ | 5.209x10$^{-5}$ | 3.671x10$^{1}$ | 4.145x10$^{1}$ | 1.958x10$^{2}$ | 7.732x10$^{0}$ | SF |
| $^{285}111$ | 8.320 | 1.037x10$^{-7}$ | 7.355x10$^{-7}$ | 1.460x10$^{5}$ | 5.144x10$^{4}$ | 2.559x10$^{5}$ | 8.605x10$^{3}$ | SF |
| $^{287}111$ | 8.170 | 3.553x10$^{-11}$ | 3.775x10$^{-9}$ | 1.237x10$^{5}$ | 1.718x10$^{5}$ | 9.237x10$^{5}$ | 2.857x10$^{4}$ | SF |
| $^{289}111$ | 8.010 | 2.357x10$^{-15}$ | 7.343x10$^{-12}$ | 7.183x10$^{5}$ | 6.513x10$^{5}$ | 3.778x10$^{6}$ | 1.078x10$^{5}$ | SF |
| $^{291}111$ | 7.600 | 3.029x10$^{-20}$ | 5.642x10$^{-15}$ | 4.002x10$^{8}$ | 2.756x10$^{7}$ | 1.706x10$^{8}$ | 4.513x10$^{6}$ | SF |
| $^{293}111$ | 7.310 | 7.544x10$^{-26}$ | 1.781x10$^{-18}$ | 8.186x10$^{9}$ | 4.614x10$^{8}$ | 3.060x10$^{9}$ | 7.591x10$^{7}$ | SF |
| $^{295}111$ | 7.200 | 3.642x10$^{-32}$ | 2.398x10$^{-22}$ | 2.594x10$^{10}$ | 1.338x10$^{9}$ | 9.571x10$^{9}$ | 2.216x10$^{8}$ | SF |
| $^{297}111$ | 8.640 | 3.411x10$^{-39}$ | 1.429x10$^{-26}$ | 1.907x10$^{4}$ | 2.255x10$^{3}$ | 1.853x10$^{4}$ | 4.048x10$^{2}$ | SF |

**Table 4.** The alpha decay half lives and the spontaneous fission half lives of $^{257-301}$113 isotopes. The mode of decay is predicted by comparing the alpha decay half lives with the spontaneous fission half lives. The alpha half lives calculations are done for zero angular momentum transfers.

| Parent nuclei | $Q_\alpha$ (cal. MeV) | $T_{SF}^{av}$ (Xu) (s) | $T_{SF}^{av}$ (KPS) (s) | $T_{1/2}^\alpha$ (s) CPPMDN | $T_{1/2}^\alpha$ (s) Royer | $T_{1/2}^\alpha$ (s) VSS | $T_{1/2}^\alpha$ (s) UNIV | Mode of Decay |
|---|---|---|---|---|---|---|---|---|
| $^{257}$113 | 13.811 | 1.714x10$^{-41}$ | 5.477x10$^{-38}$ | 1.199x10$^{-8}$ | 1.284x10$^{-8}$ | 1.750x10$^{-8}$ | 6.072x10$^{-9}$ | SF |
| $^{259}$113 | 13.681 | 3.154x10$^{-34}$ | 3.519x10$^{-31}$ | 1.924x10$^{-8}$ | 1.988x10$^{-8}$ | 2.963x10$^{-8}$ | 9.127x10$^{-9}$ | SF |
| $^{261}$113 | 13.561 | 1.112x10$^{-27}$ | 3.522x10$^{-25}$ | 2.984x10$^{-8}$ | 2.978x10$^{-8}$ | 4.850x10$^{-8}$ | 1.331x10$^{-8}$ | SF |
| $^{263}$113 | 13.301 | 7.523x10$^{-22}$ | 5.946x10$^{-20}$ | 1.011x10$^{-9}$ | 8.127x10$^{-8}$ | 1.443x10$^{-7}$ | 3.377x10$^{-8}$ | SF |
| $^{265}$113 | 13.091 | 9.774x10$^{-17}$ | 1.830x10$^{-15}$ | 2.412x10$^{-9}$ | 1.841x10$^{-7}$ | 3.563x10$^{-7}$ | 7.238x10$^{-8}$ | SF |
| $^{267}$113 | 12.951 | 2.446x10$^{-12}$ | 1.110x10$^{-11}$ | 5.685x10$^{-9}$ | 3.123x10$^{-7}$ | 6.590x10$^{-7}$ | 1.189x10$^{-7}$ | SF |
| $^{269}$113 | 12.771 | 1.195x10$^{-8}$ | 1.445x10$^{-8}$ | 2.140x10$^{-8}$ | 6.416x10$^{-7}$ | 1.475x10$^{-6}$ | 2.335x10$^{-7}$ | SF |
| $^{271}$113 | 12.601 | 1.216x10$^{-5}$ | 4.541x10$^{-6}$ | 6.127x10$^{-8}$ | 1.281x10$^{-6}$ | 3.206x10$^{-6}$ | 4.475x10$^{-7}$ | α |
| $^{273}$113 | 12.371 | 3.278x10$^{-3}$ | 4.102x10$^{-4}$ | 3.618x10$^{-7}$ | 3.456x10$^{-6}$ | 9.401x10$^{-6}$ | 1.138x10$^{-6}$ | α |
| $^{275}$113 | 12.151 | 3.239x10$^{-1}$ | 1.284x10$^{-2}$ | 1.911x10$^{-6}$ | 9.159x10$^{-6}$ | 2.707x10$^{-5}$ | 2.855x10$^{-6}$ | α |
| $^{277}$113 | 11.921 | 9.853x10$^{0}$ | 1.452x10$^{-1}$ | 2.502x10$^{-7}$ | 2.630x10$^{-5}$ | 8.436x10$^{-5}$ | 7.746x10$^{-6}$ | α |
| $^{279}$113 | 11.611 | 6.676x10$^{1}$ | 5.449x10$^{-1}$ | 6.074x10$^{-5}$ | 1.185x10$^{-4}$ | 4.118x10$^{-4}$ | 3.229x10$^{-5}$ | α |
| $^{281}$113 | 11.061 | 8.988x10$^{1}$ | 6.455x10$^{-1}$ | 5.555x10$^{-5}$ | 2.154x10$^{-3}$ | 8.066x10$^{-3}$ | 5.120x10$^{-4}$ | α |
| $^{283}$113 | 10.541 | 2.348x10$^{1}$ | 2.428x10$^{-1}$ | 2.654x10$^{-3}$ | 4.117x10$^{-2}$ | 1.660x10$^{-1}$ | 8.709x10$^{-3}$ | α |
| $^{285}$113 | 10.091 | 1.186x10$^{0}$ | 2.998x10$^{-2}$ | 5.381x10$^{-1}$ | 6.309x10$^{-1}$ | 2.740x10$^{0}$ | 1.221x10$^{-1}$ | SF |
| $^{287}$113 | 9.601 | 1.158x10$^{-2}$ | 1.268x10$^{-3}$ | 7.477x10$^{0}$ | 1.551x10$^{1}$ | 7.240x10$^{1}$ | 2.759x10$^{0}$ | SF |
| $^{289}$113 | 8.851 | 2.185x10$^{-5}$ | 1.916x10$^{-5}$ | 2.845x10$^{4}$ | 3.666x10$^{3}$ | 1.823x10$^{4}$ | 5.880x10$^{2}$ | SF |
| $^{291}$113 | 8.311 | 7.978x10$^{-9}$ | 1.080x10$^{-7}$ | 5.626x10$^{5}$ | 2.896x10$^{5}$ | 1.540x10$^{6}$ | 4.436x10$^{4}$ | SF |
| $^{293}$113 | 8.111 | 5.638x10$^{-13}$ | 2.367x10$^{-10}$ | 1.796x10$^{7}$ | 1.551x10$^{6}$ | 8.900x10$^{6}$ | 2.359x10$^{5}$ | SF |
| $^{295}$113 | 8.001 | 7.716x10$^{-18}$ | 2.097x10$^{-13}$ | 5.012x10$^{7}$ | 3.871x10$^{6}$ | 2.402x10$^{7}$ | 5.889x10$^{5}$ | SF |
| $^{297}$113 | 7.851 | 2.046x10$^{-23}$ | 7.805x10$^{-17}$ | 2.064x10$^{8}$ | 1.436x10$^{7}$ | 9.618x10$^{7}$ | 2.184x10$^{6}$ | SF |
| $^{299}$113 | 9.421 | 1.052x10$^{-29}$ | 1.265x10$^{-20}$ | 2.127x10$^{2}$ | 3.381x10$^{1}$ | 2.569x10$^{2}$ | 6.091x10$^{0}$ | SF |
| $^{301}$113 | 9.141 | 1.049x10$^{-36}$ | 9.254x10$^{-25}$ | 1.784x10$^{3}$ | 2.426x10$^{2}$ | 1.984x10$^{3}$ | 4.208x10$^{1}$ | SF |

**Table 5.** The alpha decay half lives and the spontaneous fission half lives of $^{261-305}$115 isotopes. The mode of decay is predicted by comparing the alpha decay half lives with the spontaneous fission half lives. The alpha half lives calculations are done for zero angular momentum transfers.

| Parent nuclei | $Q_\alpha$ (cal. MeV) | $T_{SF}^{av}$ (Xu) (s) | $T_{SF}^{av}$ (KPS) (s) | $T_{1/2}^\alpha$ (s) | | | | Mode of Decay |
|---|---|---|---|---|---|---|---|---|
| | | | | CPPMDN | Royer | VSS | UNIV | |
| $^{261}$115 | 17.863 | 1.389x10$^{-39}$ | 3.126x10$^{-35}$ | 5.890x10$^{-14}$ | 4.682x10$^{-14}$ | 6.359x10$^{-14}$ | 9.448x10$^{-14}$ | SF |
| $^{263}$115 | 17.013 | 2.724x10$^{-32}$ | 1.147x10$^{-28}$ | 6.694x10$^{-13}$ | 5.012x10$^{-13}$ | 7.404x10$^{-13}$ | 7.132x10$^{-13}$ | SF |
| $^{265}$115 | 14.013 | 1.024x10$^{-25}$ | 7.007x10$^{-23}$ | 1.469x10$^{-8}$ | 1.520x10$^{-8}$ | 2.389x10$^{-8}$ | 6.854x10$^{-9}$ | SF |
| $^{267}$115 | 13.803 | 7.397x10$^{-20}$ | 7.731x10$^{-18}$ | 3.370x10$^{-8}$ | 3.263x10$^{-8}$ | 5.597x10$^{-8}$ | 1.389x10$^{-8}$ | SF |
| $^{269}$115 | 13.683 | 1.032x10$^{-14}$ | 1.682x10$^{-13}$ | 5.261x10$^{-8}$ | 4.904x10$^{-8}$ | 9.183x10$^{-8}$ | 2.033x10$^{-8}$ | SF |
| $^{271}$115 | 13.543 | 2.873x10$^{-10}$ | 8.086x10$^{-10}$ | 3.201x10$^{-9}$ | 8.077x10$^{-8}$ | 1.650x10$^{-7}$ | 3.239x10$^{-8}$ | SF |
| $^{273}$115 | 13.383 | 1.830x10$^{-6}$ | 1.009x10$^{-6}$ | 8.003x10$^{-9}$ | 1.464x10$^{-7}$ | 3.260x10$^{-7}$ | 5.643x10$^{-8}$ | α |
| $^{275}$115 | 13.173 | 3.767x10$^{-3}$ | 3.908x10$^{-4}$ | 3.972x10$^{-8}$ | 3.349x10$^{-7}$ | 8.117x10$^{-7}$ | 1.221x10$^{-7}$ | α |
| $^{277}$115 | 12.983 | 2.775x10$^{0}$ | 4.999x10$^{-2}$ | 6.747x10$^{-9}$ | 7.160x10$^{-7}$ | 1.889x10$^{-6}$ | 2.486x10$^{-7}$ | α |
| $^{279}$115 | 12.783 | 5.169x10$^{2}$ | 1.996x10$^{0}$ | 5.899x10$^{-7}$ | 1.634x10$^{-6}$ | 4.689x10$^{-6}$ | 5.393x10$^{-7}$ | α |
| $^{281}$115 | 12.433 | 1.978x10$^{4}$ | 2.373x10$^{1}$ | 4.052x10$^{-6}$ | 7.795x10$^{-6}$ | 2.426x10$^{-5}$ | 2.336x10$^{-6}$ | α |
| $^{283}$115 | 11.813 | 1.477x10$^{5}$ | 8.438x10$^{1}$ | 6.674x10$^{-6}$ | 1.584x10$^{-4}$ | 5.326x10$^{-4}$ | 4.009x10$^{-5}$ | α |
| $^{285}$115 | 10.743 | 2.132x10$^{5}$ | 9.269x10$^{1}$ | 4.108x10$^{-2}$ | 5.660x10$^{-2}$ | 2.038x10$^{-1}$ | 1.102x10$^{-2}$ | α |
| $^{287}$115 | 10.793 | 5.940x10$^{4}$ | 3.282x10$^{1}$ | 3.462x10$^{-2}$ | 3.860x10$^{-2}$ | 1.514x10$^{-1}$ | 7.674x10$^{-3}$ | α |
| $^{289}$115 | 10.573 | 3.196x10$^{3}$ | 3.917x10$^{0}$ | 8.647x10$^{-2}$ | 1.340x10$^{-1}$ | 5.692x10$^{-1}$ | 2.557x10$^{-2}$ | α |
| $^{291}$115 | 10.373 | 3.323x10$^{1}$ | 1.647x10$^{-1}$ | 4.867x10$^{-1}$ | 4.276x10$^{-1}$ | 1.968x10$^{0}$ | 7.882x10$^{-2}$ | α |
| $^{293}$115 | 9.363 | 6.679x10$^{-2}$ | 2.545x10$^{-3}$ | 2.464x10$^{3}$ | 3.824x10$^{2}$ | 1.877x10$^{3}$ | 5.928x10$^{1}$ | SF |
| $^{295}$115 | 9.523 | 2.597x10$^{-5}$ | 1.507x10$^{-5}$ | 6.922x10$^{2}$ | 1.101x10$^{2}$ | 5.889x10$^{2}$ | 1.762x10$^{1}$ | SF |
| $^{297}$115 | 9.403 | 1.954x10$^{-9}$ | 3.559x10$^{-8}$ | 1.633x10$^{3}$ | 2.419x10$^{2}$ | 1.401x10$^{3}$ | 3.825x10$^{1}$ | SF |
| $^{299}$115 | 9.233 | 2.847x10$^{-14}$ | 3.479x10$^{-11}$ | 5.853x10$^{3}$ | 7.860x10$^{2}$ | 4.921x103 | 1.221x10$^{2}$ | SF |
| $^{301}$115 | 10.713 | 8.037x10$^{-20}$ | 1.460x10$^{-14}$ | 1.342x10$^{-1}$ | 3.496x10$^{-2}$ | 2.439x10$^{-1}$ | 7.259x10$^{-3}$ | SF |
| $^{303}$115 | 10.293 | 4.398x10$^{-26}$ | 2.727x10$^{-18}$ | 2.036x10$^{0}$ | 4.349x10$^{-1}$ | 3.265x10$^{0}$ | 8.280x10$^{-2}$ | SF |
| $^{305}$115 | 9.643 | 4.667x10$^{-33}$ | 2.342x10$^{-22}$ | 2.006x10$^{2}$ | 3.130x10$^{1}$ | 2.516x10$^{2}$ | 5.290x10$^{0}$ | SF |

**Table 6.** The alpha decay half lives and the spontaneous fission half lives of $^{267-309}$117 isotopes. The mode of decay is predicted by comparing the alpha decay half lives with the spontaneous fission half lives. The alpha half lives calculations are done for zero angular momentum transfers.

| Parent nuclei | $Q_\alpha$ (cal. MeV) | $T_{SF}^{av}$ (Xu) (s) | $T_{SF}^{av}$ (KPS) (s) | $T_{1/2}^\alpha$ (s) | | | | Mode of Decay |
| --- | --- | --- | --- | --- | --- | --- | --- | --- |
| | | | | CPPMDN | Royer | VSS | UNIV | |
| $^{267}$117 | 11.984 | 3.432x10$^{-29}$ | 3.448x10$^{-25}$ | 1.394x10$^{-3}$ | 5.900x10$^{-4}$ | 7.949x10$^{-4}$ | 1.232x10$^{-4}$ | SF |
| $^{269}$117 | 14.514 | 1.381x10$^{-22}$ | 1.295x10$^{-19}$ | 6.062x10$^{-9}$ | 6.578x10$^{-9}$ | 9.908x10$^{-9}$ | 3.049x10$^{-9}$ | SF |
| $^{271}$117 | 14.394 | 1.088x10$^{-16}$ | 1.005x10$^{-14}$ | 9.162x10$^{-9}$ | 9.609x10$^{-9}$ | 1.581x10$^{-8}$ | 4.336x10$^{-9}$ | SF |
| $^{273}$117 | 14.294 | 1.824x10$^{-11}$ | 1.896x10$^{-10}$ | 1.281x10$^{-8}$ | 1.305x10$^{-8}$ | 2.345x10$^{-8}$ | 5.771x10$^{-9}$ | SF |
| $^{275}$117 | 14.094 | 8.618x10$^{-7}$ | 1.014x10$^{-6}$ | 2.794x10$^{-8}$ | 2.664x10$^{-8}$ | 5.220x10$^{-8}$ | 1.115x10$^{-8}$ | α |
| $^{277}$117 | 13.924 | 1.510x10$^{-2}$ | 1.598x10$^{-3}$ | 5.853x10$^{-9}$ | 4.892x10$^{-8}$ | 1.045x10$^{-7}$ | 1.959x10$^{-8}$ | α |
| $^{279}$117 | 13.774 | 7.620x10$^{1}$ | 7.113x10$^{-1}$ | 1.077x10$^{-9}$ | 8.369x10$^{-8}$ | 1.948x10$^{-7}$ | 3.230x10$^{-8}$ | α |
| $^{281}$117 | 13.024 | 8.273x10$^{4}$ | 8.707x10$^{1}$ | 9.806x10$^{-7}$ | 2.041x10$^{-6}$ | 5.148x10$^{-6}$ | 6.247x10$^{-7}$ | α |
| $^{283}$117 | 12.564 | 1.768x10$^{7}$ | 2.976x10$^{3}$ | 3.992x10$^{-6}$ | 1.617x10$^{-5}$ | 4.428x10$^{-5}$ | 4.353x10$^{-6}$ | α |
| $^{285}$117 | 11.934 | 7.310x10$^{8}$ | 2.948x10$^{4}$ | 7.890x10$^{-5}$ | 3.473x10$^{-4}$ | 1.030x10$^{-3}$ | 7.898x10$^{-5}$ | α |
| $^{287}$117 | 11.504 | 5.828x10$^{9}$ | 8.855x10$^{4}$ | 8.521x10$^{-4}$ | 3.177x10$^{-3}$ | 1.021x10$^{-2}$ | 6.520x10$^{-4}$ | α |
| $^{289}$117 | 11.214 | 8.962x10$^{9}$ | 8.448x10$^{4}$ | 5.134x10$^{-3}$ | 1.482x10$^{-2}$ | 5.168x10$^{-2}$ | 2.858x10$^{-3}$ | α |
| $^{291}$117 | 11.444 | 2.660x10$^{9}$ | 2.681x10$^{4}$ | 4.172x10$^{-3}$ | 3.737x10$^{-3}$ | 1.421x10$^{-2}$ | 7.704x10$^{-4}$ | α |
| $^{293}$117 | 11.244 | 1.524x10$^{8}$ | 2.958x10$^{3}$ | 1.147x10$^{-2}$ | 1.056x10$^{-2}$ | 4.357x10$^{-2}$ | 2.091x10$^{-3}$ | α |
| $^{295}$117 | 10.644 | 1.687x10$^{6}$ | 1.185x10$^{2}$ | 1.409x10$^{0}$ | 3.402x10$^{-1}$ | 1.515x10$^{0}$ | 5.917x10$^{-2}$ | α |
| $^{297}$117 | 10.864 | 3.611x10$^{3}$ | 1.795x10$^{0}$ | 3.018x10$^{-1}$ | 8.218x10$^{-2}$ | 3.985x10$^{-1}$ | 1.514x10$^{-2}$ | α |
| $^{299}$117 | 10.734 | 1.494x10$^{0}$ | 1.070x10$^{-2}$ | 6.765x10$^{-1}$ | 1.660x10$^{-1}$ | 8.729x10$^{-1}$ | 2.996x10$^{-2}$ | α |
| $^{301}$117 | 10.604 | 1.197x10$^{-4}$ | 2.609x10$^{-5}$ | 1.486x10$^{0}$ | 3.405x10$^{-1}$ | 1.940x10$^{0}$ | 6.021x10$^{-2}$ | SF |
| $^{303}$117 | 12.144 | 1.856x10$^{-9}$ | 2.696x10$^{-8}$ | 1.326x10$^{-4}$ | 5.588x10$^{-5}$ | 3.511x10$^{-4}$ | 1.484x10$^{-5}$ | SF |
| $^{305}$117 | 11.484 | 5.577x10$^{-15}$ | 1.224x10$^{-11}$ | 5.257x10$^{-3}$ | 1.685x10$^{-3}$ | 1.140x10$^{-2}$ | 3.753x10$^{-4}$ | SF |
| $^{307}$117 | 10.744 | 3.247x10$^{-21}$ | 2.523x10$^{-15}$ | 4.859x10$^{-1}$ | 1.129x10$^{-1}$ | 8.214x10$^{-1}$ | 2.112x10$^{-2}$ | SF |
| $^{309}$117 | 9.884 | 3.667x10$^{-28}$ | 2.440x10$^{-19}$ | 1.722x10$^{2}$ | 2.745x10$^{1}$ | 2.140x10$^{2}$ | 4.342x10$^{0}$ | SF |

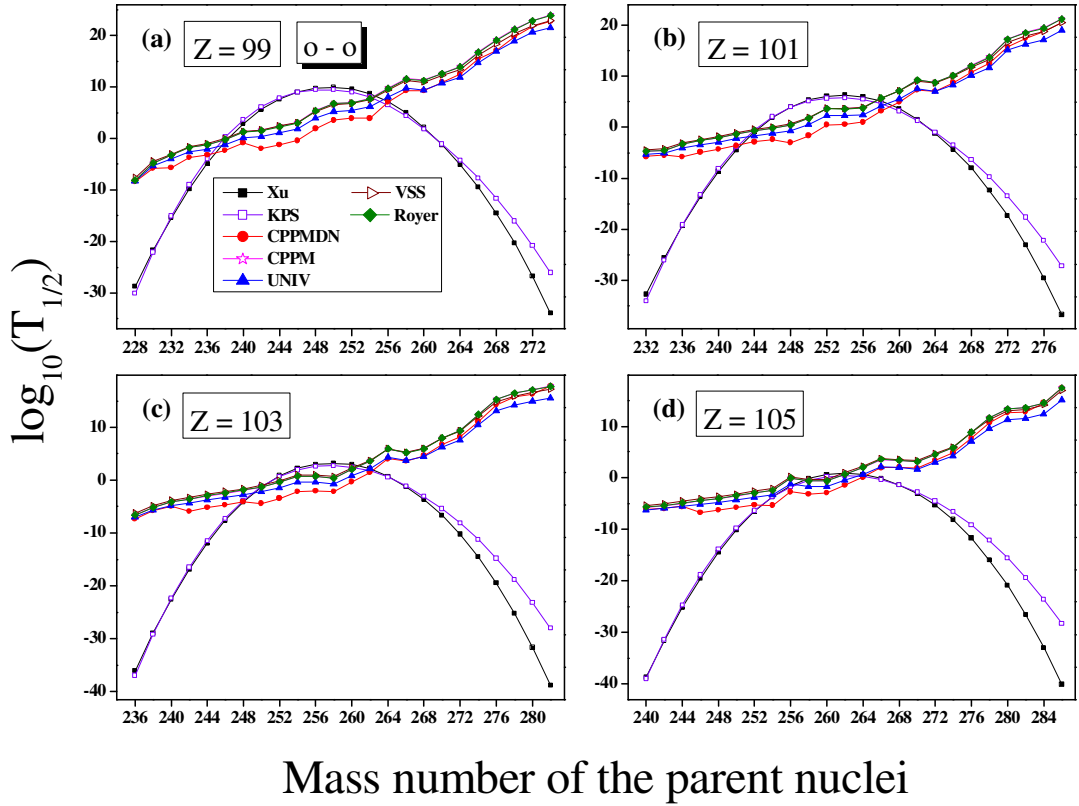

**Figure 5.** (Color online) Plot for the comparison of the computed alpha decay half lives with the spontaneous fission half lives for the odd-odd, (a) $^{228-274}$99 (b) $^{232-278}$101 (c) $^{236-282}$103 and (d) $^{240-286}$105, isotopes.

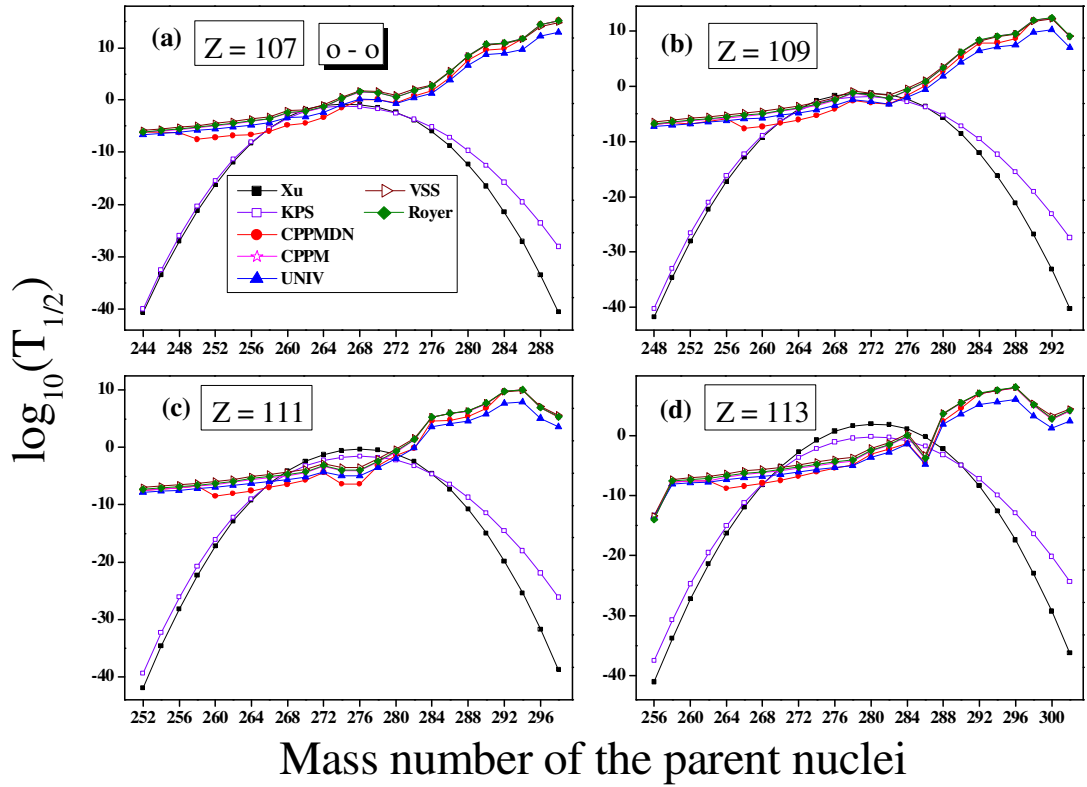

**Figure 6.** (Color online) Plot for the comparison of the computed alpha decay half lives with the spontaneous fission half lives for the odd-odd, (a) $^{244-290}$107 (b) $^{248-294}$109 (c) $^{252-298}$111 and (d) $^{256-302}$113, isotopes.

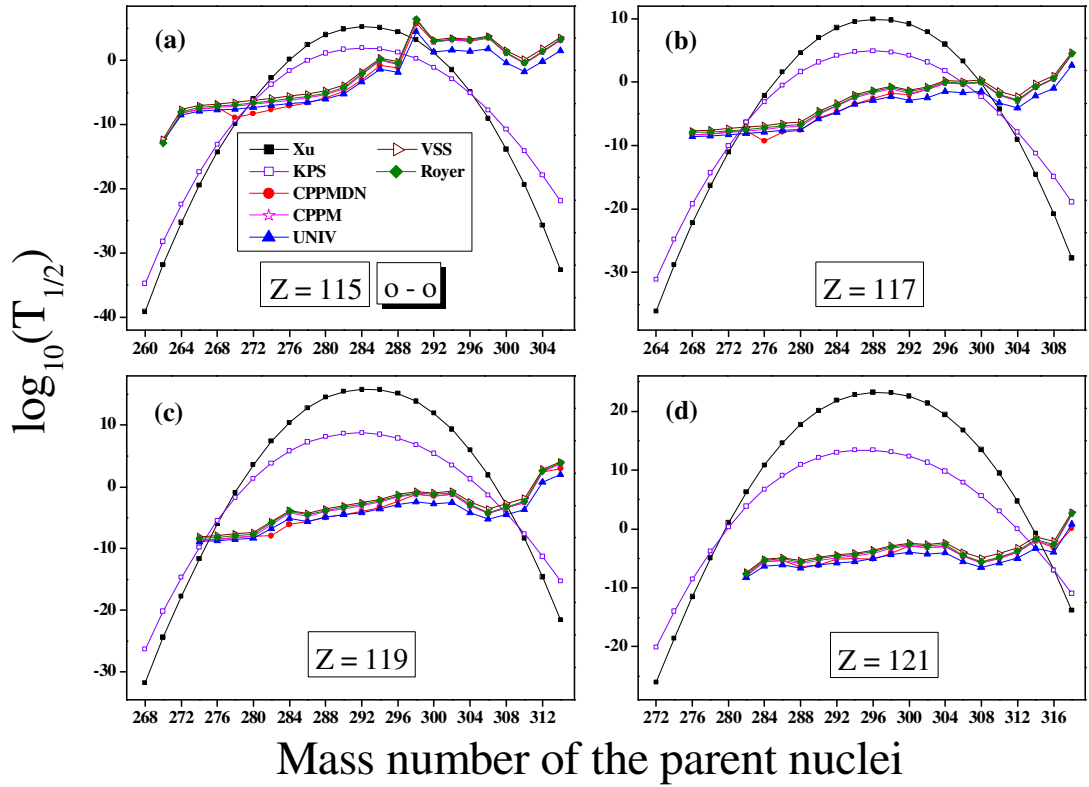

**Figure 7.** (Color online) Plot for the comparison of the computed alpha decay half lives with the spontaneous fission half lives for the odd-odd, (a) $^{260-306}$115 (b) $^{264-310}$117 (c) $^{268-314}$119 and (d) $^{272-318}$121, isotopes.

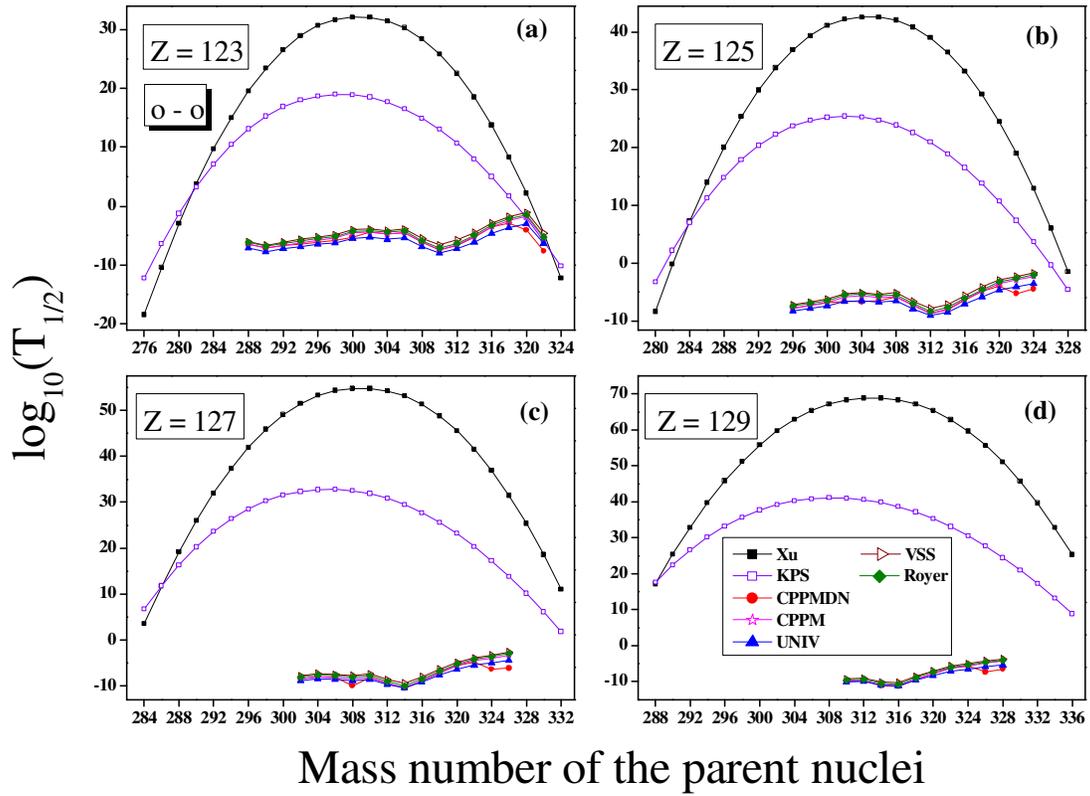

**Figure 8.** (Color online) Plot for the comparison of the computed alpha decay half lives with the spontaneous fission half lives for the odd-odd, (a) $^{276-324}$123 (b) $^{280-328}$125 (c) $^{284-332}$127 and (d) $^{288-336}$129, isotopes.

**Table 7.** A comparison of the alpha decay half lives of the experimentally synthesised odd-odd superheavy nuclei with the corresponding experimental alpha half lives [14, 18]. The alpha half lives calculations are done for zero angular momentum transfers.

| Parent Nuclei | $Q_\alpha$ (cal.) MeV | $T^\alpha_{1/2}$ (s) (Expt.) | $T^\alpha_{1/2}$ (s) | | | | |
|---|---|---|---|---|---|---|---|
| | | | CPPMDN | CPPM | VSS | UNIV | Royer |
| $^{270}$109 | 10.228 | 5.754x10$^{-1}$ | 2.431x10$^{-3}$ | 5.357x10$^{-2}$ | 1.353x10$^{-1}$ | 4.449x10$^{-3}$ | 6.588x10$^{-2}$ |
| $^{276}$109 | 10.039 | 7.200x10$^{-1}$ | 2.576x10$^{-2}$ | 1.503x10$^{-1}$ | 4.335x10$^{-1}$ | 1.093x10$^{-2}$ | 1.788x10$^{-1}$ |
| $^{278}$109 | 9.518 | 7.700x10$^{0}$ | 1.048x10$^{0}$ | 5.371x10$^{0}$ | 1.283x10$^{1}$ | 2.736x10$^{-1}$ | 6.210x10$^{0}$ |
| $^{280}$111 | 10.259 | 3.600x10$^{0}$ | 4.968x10$^{-2}$ | 1.733x10$^{-1}$ | 4.832x10$^{-1}$ | 1.151x10$^{-2}$ | 2.195x10$^{-1}$ |
| $^{282}$111 | 9.560 | 5.100x10$^{-1}$ | 7.849x10$^{-1}$ | 2.131x10$^{1}$ | 4.538x10$^{1}$ | 8.862x10$^{-1}$ | 2.619x10$^{1}$ |
| $^{284}$113 | 10.281 | 4.786x10$^{-1}$ | 4.332x10$^{-2}$ | 7.427x10$^{-1}$ | 1.802x10$^{0}$ | 3.933x10$^{-2}$ | 9.883x10$^{-1}$ |
| $^{288}$115 | 10.693 | 8.700x10$^{-2}$ | 7.031x10$^{-2}$ | 2.389x10$^{-1}$ | 6.043x10$^{-1}$ | 1.315x10$^{-2}$ | 3.404x10$^{-1}$ |
| $^{294}$117 | 11.144 | 7.800x10$^{-2}$ | 6.773x10$^{-2}$ | 6.773x10$^{-2}$ | 1.903x10$^{-1}$ | 3.894x10$^{-3}$ | 1.020x10$^{-1}$ |

**Table 8.** The alpha decay half lives and the spontaneous fission half lives of $^{248-294}$109 isotopes. The mode of decay is predicted by comparing the alpha decay half lives with the spontaneous fission half lives. The alpha half lives calculations are done for zero angular momentum transfers.

| Parent nuclei | $Q_\alpha$ (cal. MeV) | $T_{SF}^{av}$ (Xu) (s) | $T_{SF}^{av}$ (KPS) (s) | $T_{1/2}^\alpha$ (s) | | | | Mode of Decay |
|---|---|---|---|---|---|---|---|---|
| | | | | CPPMDN | Royer | VSS | UNIV | |
| $^{248}$109 | 12.718 | 1.860x10$^{-42}$ | 6.571x10$^{-41}$ | 1.310x10$^{-7}$ | 1.837x10$^{-7}$ | 3.634x10$^{-7}$ | 5.296x10$^{-8}$ | SF |
| $^{250}$109 | 12.558 | 3.009x10$^{-35}$ | 1.167x10$^{-33}$ | 2.559x10$^{-7}$ | 3.585x10$^{-7}$ | 7.375x10$^{-7}$ | 9.475x10$^{-8}$ | SF |
| $^{252}$109 | 12.398 | 9.325x10$^{-29}$ | 2.819x10$^{-27}$ | 1.687x10$^{-7}$ | 7.104x10$^{-7}$ | 1.517x10$^{-6}$ | 1.720x10$^{-7}$ | SF |
| $^{254}$109 | 12.238 | 5.544x10$^{-23}$ | 1.011x10$^{-21}$ | 4.816x10$^{-7}$ | 1.430x10$^{-6}$ | 3.166x10$^{-6}$ | 3.170x10$^{-7}$ | SF |
| $^{256}$109 | 12.098 | 6.327x10$^{-18}$ | 5.853x10$^{-17}$ | 1.481x10$^{-6}$ | 2.643x10$^{-6}$ | 6.097x10$^{-6}$ | 5.429x10$^{-7}$ | SF |
| $^{258}$109 | 11.918 | 1.387x10$^{-13}$ | 5.926x10$^{-13}$ | 2.029x10$^{-8}$ | 6.084x10$^{-6}$ | 1.440x10$^{-5}$ | 1.129x10$^{-6}$ | SF |
| $^{260}$109 | 11.778 | 5.849x10$^{-10}$ | 1.133x10$^{-9}$ | 5.129x10$^{-8}$ | 1.159x10$^{-5}$ | 2.849x10$^{-5}$ | 1.990x10$^{-6}$ | SF |
| $^{262}$109 | 11.538 | 4.752x10$^{-7}$ | 4.408x10$^{-7}$ | 2.189x10$^{-7}$ | 3.838x10$^{-5}$ | 9.444x10$^{-5}$ | 5.727x10$^{-6}$ | α |
| $^{264}$109 | 11.318 | 7.499x10$^{-5}$ | 3.779x10$^{-5}$ | 8.808x10$^{-7}$ | 1.185x10$^{-4}$ | 2.928x10$^{-4}$ | 1.554x10$^{-5}$ | α |
| $^{266}$109 | 11.048 | 2.385x10$^{-3}$ | 7.903x10$^{-4}$ | 5.265x10$^{-6}$ | 5.062x10$^{-4}$ | 1.230x10$^{-3}$ | 5.652x10$^{-5}$ | α |
| $^{268}$109 | 10.718 | 1.793x10$^{-2}$ | 4.791x10$^{-3}$ | 6.376x10$^{-5}$ | 3.294x10$^{-3}$ | 7.641x10$^{-3}$ | 3.011x10$^{-4}$ | α |
| $^{270}$109 | 10.228 | 4.686x10$^{-2}$ | 1.102x10$^{-2}$ | 2.431x10$^{-3}$ | 6.588x10$^{-2}$ | 1.353x10$^{-1}$ | 4.449x10$^{-3}$ | α |
| $^{272}$109 | 10.398 | 5.440x10$^{-2}$ | 1.185x10$^{-2}$ | 1.042x10$^{-3}$ | 2.031x10$^{-2}$ | 4.879x10$^{-2}$ | 1.539x10$^{-3}$ | α |
| $^{274}$109 | 10.558 | 2.791x10$^{-2}$ | 6.284x10$^{-3}$ | 6.351x10$^{-4}$ | 6.842x10$^{-3}$ | 1.911x10$^{-2}$ | 5.784x10$^{-4}$ | α |
| $^{276}$109 | 10.039 | 4.952x10$^{-3}$ | 1.566x10$^{-3}$ | 2.576x10$^{-2}$ | 1.788x10$^{-1}$ | 4.335x10$^{-1}$ | 1.093x10$^{-2}$ | SF |
| $^{278}$109 | 9.518 | 2.081x10$^{-4}$ | 1.564x10$^{-4}$ | 1.048x10$^{0}$ | 6.210x10$^{0}$ | 1.283x10$^{1}$ | 2.736x10$^{-1}$ | SF |
| $^{280}$109 | 8.748 | 1.769x10$^{-6}$ | 5.577x10$^{-6}$ | 4.354x10$^{2}$ | 2.191x10$^{3}$ | 3.299x10$^{3}$ | 5.907x10$^{1}$ | SF |
| $^{282}$109 | 8.008 | 2.935x10$^{-9}$ | 6.936x10$^{-8}$ | 1.982x10$^{5}$ | 1.352x10$^{6}$ | 1.430x10$^{6}$ | 2.261x10$^{4}$ | SF |
| $^{284}$109 | 7.498 | 9.438x10$^{-13}$ | 3.080x10$^{-10}$ | 5.512x10$^{7}$ | 1.916x10$^{8}$ | 1.568x10$^{8}$ | 2.298x10$^{6}$ | SF |
| $^{286}$109 | 7.328 | 5.877x10$^{-17}$ | 5.071x10$^{-13}$ | 6.888x10$^{7}$ | 1.063x10$^{9}$ | 8.364x10$^{8}$ | 1.142x10$^{7}$ | SF |
| $^{288}$109 | 7.218 | 7.089x10$^{-22}$ | 3.223x10$^{-16}$ | 3.744x10$^{8}$ | 3.241x10$^{9}$ | 2.549x10$^{9}$ | 3.244x10$^{7}$ | SF |
| $^{290}$109 | 6.728 | 1.657x10$^{-27}$ | 8.236x10$^{-20}$ | 8.092x10$^{11}$ | 8.696x10$^{11}$ | 5.067x10$^{11}$ | 6.273x10$^{9}$ | SF |
| $^{292}$109 | 6.648 | 7.513x10$^{-34}$ | 8.797x10$^{-24}$ | 1.966x10$^{12}$ | 2.152x10$^{12}$ | 1.270x10$^{12}$ | 1.473x10$^{10}$ | SF |
| $^{294}$109 | 7.308 | 6.608x10$^{-41}$ | 4.079x10$^{-28}$ | 9.286x10$^{8}$ | 9.609x10$^{8}$ | 1.022x10$^{9}$ | 1.030x10$^{7}$ | SF |

**Table 9.** The alpha decay half lives and the spontaneous fission half lives of $^{252-298}111$ isotopes. The mode of decay is predicted by comparing the alpha decay half lives with the spontaneous fission half lives. The alpha half lives calculations are done for zero angular momentum transfers.

| Parent nuclei | $Q_\alpha$ (cal. MeV) | $T_{SF}^{av}$ (Xu) (s) | $T_{SF}^{av}$ (KPS) (s) | $T_{1/2}^{\alpha}$ (s) | | | | Mode of Decay |
|---|---|---|---|---|---|---|---|---|
| | | | | CPPMDN | Royer | VSS | UNIV | |
| $^{252}111$ | 13.290 | 1.269x10$^{-42}$ | 4.435x10$^{-40}$ | 3.499x10$^{-8}$ | 5.287x10$^{-8}$ | 1.036x10$^{-7}$ | 1.604x10$^{-8}$ | SF |
| $^{254}111$ | 13.170 | 2.189x10$^{-35}$ | 4.826x10$^{-33}$ | 5.480x10$^{-8}$ | 8.303x10$^{-8}$ | 1.717x10$^{-7}$ | 2.369x10$^{-8}$ | SF |
| $^{256}111$ | 13.030 | 7.239x10$^{-29}$ | 7.650x10$^{-27}$ | 9.519x10$^{-8}$ | 1.441x10$^{-7}$ | 3.122x10$^{-7}$ | 3.819x10$^{-8}$ | SF |
| $^{258}111$ | 12.870 | 4.591x10$^{-23}$ | 1.920x10$^{-21}$ | 1.843x10$^{-7}$ | 2.778x10$^{-7}$ | 6.259x10$^{-7}$ | 6.749x10$^{-8}$ | SF |
| $^{260}111$ | 12.720 | 5.590x10$^{-18}$ | 8.268x10$^{-17}$ | 3.012x10$^{-9}$ | 5.181x10$^{-7}$ | 1.216x10$^{-6}$ | 1.160x10$^{-7}$ | SF |
| $^{262}111$ | 12.530 | 1.308x10$^{-13}$ | 6.592x10$^{-13}$ | 8.986x10$^{-9}$ | 1.190x10$^{-6}$ | 2.867x10$^{-6}$ | 2.395x10$^{-7}$ | SF |
| $^{264}111$ | 12.290 | 5.892x10$^{-10}$ | 1.049x10$^{-9}$ | 2.663x10$^{-8}$ | 3.589x10$^{-6}$ | 8.718x10$^{-6}$ | 6.290x10$^{-7}$ | SF |
| $^{266}111$ | 12.120 | 5.149x10$^{-7}$ | 3.606x10$^{-7}$ | 1.024x10$^{-7}$ | 7.819x10$^{-6}$ | 1.955x10$^{-5}$ | 1.245x10$^{-6}$ | α |
| $^{268}111$ | 11.920 | 9.057x10$^{-5}$ | 2.948x10$^{-5}$ | 3.604x10$^{-7}$ | 2.033x10$^{-5}$ | 5.169x10$^{-5}$ | 2.886x10$^{-6}$ | α |
| $^{270}111$ | 11.710 | 3.763x10$^{-3}$ | 6.684x10$^{-4}$ | 1.926x10$^{-6}$ | 5.731x10$^{-5}$ | 1.473x10$^{-4}$ | 7.204x10$^{-6}$ | α |
| $^{272}111$ | 11.250 | 5.319x10$^{-2}$ | 5.322x10$^{-3}$ | 4.217x10$^{-5}$ | 6.851x10$^{-4}$ | 1.618x10$^{-3}$ | 6.512x10$^{-5}$ | α |
| $^{274}111$ | 11.540 | 2.853x10$^{-1}$ | 1.812x10$^{-2}$ | 4.732x10$^{-7}$ | 1.228x10$^{-4}$ | 3.513x10$^{-4}$ | 1.413x10$^{-5}$ | α |
| $^{276}111$ | 11.540 | 5.148x10$^{-1}$ | 2.795x10$^{-2}$ | 4.382x10$^{-7}$ | 1.129x10$^{-4}$ | 3.513x10$^{-4}$ | 1.311x10$^{-5}$ | α |
| $^{278}111$ | 10.900 | 3.466x10$^{-1}$ | 1.983x10$^{-2}$ | 7.335x10$^{-4}$ | 4.184x10$^{-3}$ | 1.108x10$^{-2}$ | 3.270x10$^{-4}$ | α |
| $^{280}111$ | 10.259 | 7.076x10$^{-2}$ | 5.965x10$^{-3}$ | 4.968x10$^{-2}$ | 2.195x10$^{-1}$ | 4.832x10$^{-1}$ | 1.151x10$^{-2}$ | SF |
| $^{282}111$ | 9.560 | 3.224x10$^{-3}$ | 6.687x10$^{-4}$ | 7.849x10$^{-1}$ | 2.619x10$^{1}$ | 4.538x10$^{1}$ | 8.862x10$^{-1}$ | SF |
| $^{284}111$ | 8.430 | 2.931x10$^{-5}$ | 2.641x10$^{-5}$ | 4.209x10$^{4}$ | 2.177x10$^{5}$ | 2.243x10$^{5}$ | 3.588x10$^{3}$ | SF |
| $^{286}111$ | 8.250 | 5.185x10$^{-8}$ | 3.696x10$^{-7}$ | 5.971x10$^{4}$ | 1.015x10$^{6}$ | 1.019x10$^{6}$ | 1.496x10$^{4}$ | SF |
| $^{288}111$ | 8.140 | 1.777x10$^{-11}$ | 1.891x10$^{-9}$ | 2.339x10$^{5}$ | 2.590x10$^{6}$ | 2.635x10$^{6}$ | 3.566x10$^{4}$ | SF |
| $^{290}111$ | 7.810 | 1.179x10$^{-15}$ | 3.675x10$^{-12}$ | 7.528x10$^{6}$ | 5.743x10$^{7}$ | 5.131x10$^{7}$ | 6.397x10$^{5}$ | SF |
| $^{292}111$ | 7.340 | 1.515x10$^{-20}$ | 2.822x10$^{-15}$ | 6.096x10$^{9}$ | 7.067x10$^{9}$ | 4.950x10$^{9}$ | 5.795x10$^{7}$ | SF |
| $^{294}111$ | 7.290 | 3.772x10$^{-26}$ | 8.905x10$^{-19}$ | 9.870x10$^{9}$ | 1.130x10$^{10}$ | 8.259x10$^{9}$ | 8.984x10$^{7}$ | SF |
| $^{296}111$ | 7.960 | 1.821x10$^{-32}$ | 1.199x10$^{-22}$ | 9.386x10$^{6}$ | 1.042x10$^{7}$ | 1.301x10$^{7}$ | 1.293x10$^{5}$ | SF |
| $^{298}111$ | 8.350 | 1.706x10$^{-39}$ | 7.145x10$^{-27}$ | 2.222x10$^{5}$ | 2.546x10$^{5}$ | 4.369x10$^{5}$ | 4.095x10$^{3}$ | SF |

**Table 10.** The alpha decay half lives and the spontaneous fission half lives of $^{256-302}$113 isotopes. The mode of decay is predicted by comparing the alpha decay half lives with the spontaneous fission half lives. The alpha half lives calculations are done for zero angular momentum transfers.

| Parent nuclei | $Q_\alpha$ (cal. MeV) | $T_{SF}^{av}$ (Xu) (s) | $T_{SF}^{av}$ (KPS) (s) | $T_{1/2}^\alpha$ (s) | | | | Mode of Decay |
| --- | --- | --- | --- | --- | --- | --- | --- | --- |
| | | | | CPPMDN | Royer | VSS | UNIV | |
| $^{256}$113 | 17.981 | 8.570x10$^{-42}$ | 2.738x10$^{-38}$ | 1.720x10$^{-14}$ | 8.540x10$^{-15}$ | 4.204x10$^{-14}$ | 3.524x10$^{-14}$ | SF |
| $^{258}$113 | 13.741 | 1.577x10$^{-34}$ | 1.759x10$^{-31}$ | 1.551x10$^{-8}$ | 2.521x10$^{-8}$ | 5.103x10$^{-8}$ | 7.576x10$^{-9}$ | SF |
| $^{260}$113 | 13.601 | 5.561x10$^{-28}$ | 1.761x10$^{-25}$ | 2.617x10$^{-8}$ | 4.253x10$^{-8}$ | 9.038x10$^{-8}$ | 1.188x10$^{-8}$ | SF |
| $^{262}$113 | 13.481 | 3.762x10$^{-22}$ | 2.973x10$^{-20}$ | 2.617x10$^{-8}$ | 6.629x10$^{-8}$ | 1.486x10$^{-7}$ | 1.741x10$^{-8}$ | SF |
| $^{264}$113 | 13.221 | 4.887x10$^{-17}$ | 9.150x10$^{-16}$ | 1.424x10$^{-9}$ | 1.970x10$^{-7}$ | 4.463x10$^{-7}$ | 4.463x10$^{-8}$ | SF |
| $^{266}$113 | 12.991 | 1.223x10$^{-12}$ | 5.549x10$^{-12}$ | 3.524x10$^{-9}$ | 5.268x10$^{-7}$ | 1.214x10$^{-6}$ | 1.047x10$^{-7}$ | SF |
| $^{268}$113 | 12.861 | 5.977x10$^{-9}$ | 7.231x10$^{-9}$ | 1.057x10$^{-8}$ | 8.962x10$^{-7}$ | 2.162x10$^{-6}$ | 1.662x10$^{-7}$ | SF |
| $^{270}$113 | 12.701 | 6.087x10$^{-6}$ | 2.278x10$^{-6}$ | 3.059x10$^{-8}$ | 1.782x10$^{-6}$ | 4.454x10$^{-6}$ | 3.026x10$^{-7}$ | α |
| $^{272}$113 | 12.471 | 1.645x10$^{-3}$ | 2.074x10$^{-4}$ | 1.593x10$^{-7}$ | 5.103x10$^{-6}$ | 1.289x10$^{-5}$ | 7.592x10$^{-7}$ | α |
| $^{274}$113 | 12.241 | 1.636x10$^{-1}$ | 6.627x10$^{-3}$ | 8.691x10$^{-7}$ | 1.509x10$^{-5}$ | 3.846x10$^{-5}$ | 1.966x10$^{-6}$ | α |
| $^{276}$113 | 12.041 | 5.089x10$^{0}$ | 7.900x10$^{-2}$ | 3.305x10$^{-6}$ | 3.940x10$^{-5}$ | 1.020x10$^{-4}$ | 4.574x10$^{-6}$ | α |
| $^{278}$113 | 11.901 | 3.831x10$^{1}$ | 3.450x10$^{-1}$ | 1.251x10$^{-5}$ | 7.643x10$^{-5}$ | 2.050x10$^{-4}$ | 8.206x10$^{-6}$ | α |
| $^{280}$113 | 11.221 | 7.832x10$^{1}$ | 5.952x10$^{-1}$ | 7.131x10$^{-4}$ | 3.211x10$^{-3}$ | 7.295x10$^{-3}$ | 2.273x10$^{-4}$ | α |
| $^{282}$113 | 10.841 | 5.668x10$^{1}$ | 4.441x10$^{-1}$ | 4.884x10$^{-3}$ | 2.920x10$^{-2}$ | 6.206x10$^{-2}$ | 1.641x10$^{-3}$ | α |
| $^{284}$113 | 10.281 | 1.234x10$^{1}$ | 1.364x10$^{-1}$ | 4.332x10$^{-2}$ | 9.883x10$^{-1}$ | 1.802x10$^{0}$ | 3.933x10$^{-2}$ | α |
| $^{286}$113 | 11.711 | 5.989x10$^{-1}$ | 1.563x10$^{-2}$ | 2.317x10$^{-5}$ | 1.546x10$^{-4}$ | 5.390x10$^{-4}$ | 1.526x10$^{-5}$ | α |
| $^{288}$113 | 9.131 | 5.800x10$^{-3}$ | 6.434x10$^{-4}$ | 2.719x10$^{2}$ | 3.789x10$^{3}$ | 4.698x10$^{3}$ | 7.369x10$^{1}$ | SF |
| $^{290}$113 | 8.611 | 1.093x10$^{-5}$ | 9.633x10$^{-6}$ | 3.447x10$^{4}$ | 2.701x10$^{5}$ | 2.734x10$^{5}$ | 3.786x10$^{3}$ | SF |
| $^{292}$113 | 8.191 | 3.989x10$^{-9}$ | 5.412x10$^{-8}$ | 9.002x10$^{6}$ | 1.125x10$^{7}$ | 9.622x10$^{6}$ | 1.208x10$^{5}$ | SF |
| $^{294}$113 | 8.061 | 2.819x10$^{-13}$ | 1.184x10$^{-10}$ | 2.899x10$^{7}$ | 3.585x10$^{7}$ | 3.064x10$^{7}$ | 3.554x10$^{5}$ | SF |
| $^{296}$113 | 7.941 | 3.858x10$^{-18}$ | 1.049x10$^{-13}$ | 8.725x10$^{7}$ | 1.068x10$^{8}$ | 9.152x10$^{7}$ | 9.822x10$^{5}$ | SF |
| $^{298}$113 | 8.661 | 1.023x10$^{-23}$ | 3.903x10$^{-17}$ | 1.062x10$^{5}$ | 1.271x10$^{5}$ | 1.821x10$^{5}$ | 1.870x10$^{3}$ | SF |
| $^{300}$113 | 9.291 | 5.259x10$^{-30}$ | 6.327x10$^{-21}$ | 5.628x10$^{2}$ | 6.641x10$^{2}$ | 1.442x10$^{3}$ | 1.472x10$^{1}$ | SF |
| $^{302}$113 | 8.911 | 5.244x10$^{-37}$ | 4.627x10$^{-25}$ | 1.126x10$^{4}$ | 1.305x10$^{4}$ | 2.510x10$^{4}$ | 2.275x10$^{2}$ | SF |

**Table 11.** The alpha decay half lives and the spontaneous fission half lives of $^{262-306}$115 isotopes. The mode of decay is predicted by comparing the alpha decay half lives with the spontaneous fission half lives. The alpha half lives calculations are done for zero angular momentum transfers.

| Parent nuclei | $Q_\alpha$ (cal. MeV) | $T_{SF}^{av}$ (Xu) (s) | $T_{SF}^{av}$ (KPS) (s) | $T_{1/2}^\alpha$ (s) | | | | Mode of Decay |
|---|---|---|---|---|---|---|---|---|
| | | | | CPPMDN | Royer | VSS | UNIV | |
| $^{262}$115 | 17.423 | 1.362x10$^{-32}$ | 5.734x10$^{-29}$ | 2.028x10$^{-13}$ | 1.215x10$^{-13}$ | 4.870x10$^{-13}$ | 2.629x10$^{-13}$ | SF |
| $^{264}$115 | 14.233 | 5.122x10$^{-26}$ | 3.503x10$^{-23}$ | 5.969x10$^{-9}$ | 1.051x10$^{-8}$ | 2.197x10$^{-8}$ | 3.207x10$^{-9}$ | SF |
| $^{266}$115 | 13.863 | 3.699x10$^{-20}$ | 3.866x10$^{-18}$ | 2.707x10$^{-8}$ | 4.670x10$^{-8}$ | 9.627x10$^{-8}$ | 1.151x10$^{-8}$ | SF |
| $^{268}$115 | 13.723 | 5.162x10$^{-15}$ | 8.412x10$^{-14}$ | 4.602x10$^{-8}$ | 7.910x10$^{-8}$ | 1.710x10$^{-7}$ | 1.812x10$^{-8}$ | SF |
| $^{270}$115 | 13.613 | 1.436x10$^{-10}$ | 4.044x10$^{-10}$ | 1.357x10$^{-9}$ | 1.183x10$^{-7}$ | 2.702x10$^{-7}$ | 2.563x10$^{-8}$ | SF |
| $^{272}$115 | 13.443 | 9.150x10$^{-7}$ | 5.051x10$^{-7}$ | 4.957x10$^{-9}$ | 2.339x10$^{-7}$ | 5.542x10$^{-7}$ | 4.618x10$^{-8}$ | α |
| $^{274}$115 | 13.263 | 1.885x10$^{-3}$ | 1.959x10$^{-4}$ | 1.901x10$^{-8}$ | 4.920x10$^{-7}$ | 1.203x10$^{-6}$ | 8.792x10$^{-8}$ | α |
| $^{276}$115 | 13.073 | 1.389x10$^{0}$ | 2.519x10$^{-2}$ | 7.885x10$^{-8}$ | 1.104x10$^{-6}$ | 2.776x10$^{-6}$ | 1.774x10$^{-7}$ | α |
| $^{278}$115 | 12.903 | 2.598x10$^{2}$ | 1.023x10$^{0}$ | 2.371x10$^{-7}$ | 2.296x10$^{-6}$ | 5.957x10$^{-6}$ | 3.355x10$^{-7}$ | α |
| $^{280}$115 | 12.643 | 1.015x10$^{4}$ | 1.286x10$^{1}$ | 1.453x10$^{-6}$ | 7.591x10$^{-6}$ | 1.973x10$^{-5}$ | 9.539x10$^{-7}$ | α |
| $^{282}$115 | 12.223 | 8.373x10$^{4}$ | 5.405x10$^{1}$ | 1.377x10$^{-5}$ | 6.013x10$^{-5}$ | 1.479x10$^{-4}$ | 5.878x10$^{-6}$ | α |
| $^{284}$115 | 11.313 | 1.804x10$^{5}$ | 8.853x10$^{1}$ | 8.386x10$^{-4}$ | 8.808x10$^{-3}$ | 1.697x10$^{-2}$ | 4.933x10$^{-4}$ | α |
| $^{286}$115 | 10.483 | 1.363x10$^{5}$ | 6.275x10$^{1}$ | 1.580x10$^{-1}$ | 1.454x10$^{0}$ | 2.177x10$^{0}$ | 4.891x10$^{-2}$ | α |
| $^{288}$115 | 10.693 | 3.130x10$^{4}$ | 1.837x10$^{1}$ | 7.031x10$^{-2}$ | 3.404x10$^{-1}$ | 6.043x10$^{-1}$ | 1.315x10$^{-2}$ | α |
| $^{290}$115 | 8.573 | 1.615x10$^{3}$ | 2.041x10$^{0}$ | 5.667x10$^{5}$ | 2.945x10$^{6}$ | 2.015x10$^{6}$ | 3.021x10$^{4}$ | SF |
| $^{292}$115 | 9.513 | 1.665x10$^{1}$ | 8.361x10$^{-2}$ | 8.310x10$^{2}$ | 1.135x10$^{3}$ | 1.390x10$^{3}$ | 2.116x10$^{1}$ | SF |
| $^{294}$115 | 9.413 | 3.341x10$^{-2}$ | 1.280x10$^{-3}$ | 1.676x10$^{3}$ | 2.265x10$^{3}$ | 2.862x10$^{3}$ | 3.986x10$^{1}$ | SF |
| $^{296}$115 | 9.463 | 1.299x10$^{-5}$ | 7.555x10$^{-6}$ | 1.061x10$^{3}$ | 1.419x10$^{3}$ | 1.992x10$^{3}$ | 2.590x10$^{1}$ | SF |
| $^{298}$115 | 9.313 | 9.770x10$^{-10}$ | 1.781x10$^{-8}$ | 3.202x10$^{3}$ | 4.230x10$^{3}$ | 5.962x10$^{3}$ | 7.053x10$^{1}$ | SF |
| $^{300}$115 | 10.043 | 1.424x10$^{-14}$ | 1.740x10$^{-11}$ | 1.274x10$^{1}$ | 1.674x10$^{1}$ | 3.630x10$^{1}$ | 4.457x10$^{-1}$ | SF |
| $^{302}$115 | 10.553 | 4.019x10$^{-20}$ | 7.303x10$^{-15}$ | 3.681x10$^{-1}$ | 4.819x10$^{-1}$ | 1.414x10$^{0}$ | 1.784x10$^{-2}$ | SF |
| $^{304}$115 | 9.953 | 2.199x10$^{-26}$ | 1.363x10$^{-18}$ | 2.128x10$^{1}$ | 2.712x10$^{1}$ | 6.604x10$^{1}$ | 6.896x10$^{-1}$ | SF |
| $^{306}$115 | 9.373 | 2.334x10$^{-33}$ | 1.171x10$^{-22}$ | 1.543x10$^{3}$ | 1.927x10$^{3}$ | 3.834x10$^{3}$ | 3.400x10$^{1}$ | SF |

**Table 12.** The alpha decay half lives and the spontaneous fission half lives of $^{268-310}117$ isotopes. The mode of decay is predicted by comparing the alpha decay half lives with the spontaneous fission half lives. The alpha half lives calculations are done for zero angular momentum transfers.

| Parent nuclei | $Q_\alpha$ (cal. MeV) | $T_{SF}^{av}$ (Xu) (s) | $T_{SF}^{av}$ (KPS) (s) | $T_{1/2}^\alpha$ (s) | | | | Mode of Decay |
|---|---|---|---|---|---|---|---|---|
| | | | | CPPMDN | Royer | VSS | UNIV | |
| $^{268}117$ | 14.524 | 6.906x10$^{-23}$ | 6.476x10$^{-20}$ | 6.103x10$^{-9}$ | 1.127x10$^{-8}$ | 2.095x10$^{-8}$ | 3.055x10$^{-9}$ | SF |
| $^{270}117$ | 14.464 | 5.440x10$^{-17}$ | 5.024x10$^{-15}$ | 7.136x10$^{-9}$ | 1.325x10$^{-8}$ | 2.644x10$^{-8}$ | 3.507x10$^{-9}$ | SF |
| $^{272}117$ | 14.324 | 9.122x10$^{-12}$ | 9.480x10$^{-11}$ | 1.180x10$^{-8}$ | 2.182x10$^{-8}$ | 4.577x10$^{-8}$ | 5.372x10$^{-9}$ | SF |
| $^{274}117$ | 14.184 | 4.309x10$^{-7}$ | 5.069x10$^{-7}$ | 1.971x10$^{-8}$ | 3.629x10$^{-8}$ | 7.986x10$^{-8}$ | 8.303x10$^{-9}$ | α |
| $^{276}117$ | 14.024 | 7.551x10$^{-3}$ | 7.996x10$^{-4}$ | 5.168x10$^{-10}$ | 6.644x10$^{-8}$ | 1.525x10$^{-7}$ | 1.395x10$^{-8}$ | α |
| $^{278}117$ | 13.834 | 3.811x10$^{1}$ | 3.564x10$^{-1}$ | 1.278x10$^{-8}$ | 1.407x10$^{-7}$ | 3.333x10$^{-7}$ | 2.660x10$^{-8}$ | α |
| $^{280}117$ | 13.734 | 4.140x10$^{4}$ | 4.389x10$^{1}$ | 2.682x10$^{-8}$ | 2.021x10$^{-7}$ | 5.064x10$^{-7}$ | 3.635x10$^{-8}$ | α |
| $^{282}117$ | 12.814 | 8.883x10$^{6}$ | 1.532x10$^{3}$ | 2.753x10$^{-6}$ | 1.442x10$^{-5}$ | 2.979x10$^{-5}$ | 1.488x10$^{-6}$ | α |
| $^{284}117$ | 12.254 | 3.743x10$^{8}$ | 1.623x10$^{4}$ | 1.952x10$^{-5}$ | 2.376x10$^{-4}$ | 4.440x10$^{-4}$ | 1.753x10$^{-5}$ | α |
| $^{286}117$ | 11.654 | 3.279x10$^{9}$ | 5.902x10$^{4}$ | 3.704x10$^{-4}$ | 6.046x10$^{-3}$ | 9.938x10$^{-3}$ | 3.110x10$^{-4}$ | α |
| $^{288}117$ | 11.344 | 7.395x10$^{9}$ | 8.651x10$^{4}$ | 2.388x10$^{-3}$ | 3.427x10$^{-2}$ | 5.449x10$^{-2}$ | 1.467x10$^{-3}$ | α |
| $^{290}117$ | 11.104 | 5.811x10$^{9}$ | 5.564x10$^{4}$ | 1.445x10$^{-2}$ | 1.358x10$^{-1}$ | 2.136x10$^{-1}$ | 5.049x10$^{-3}$ | α |
| $^{292}117$ | 11.344 | 1.406x10$^{9}$ | 1.488x10$^{4}$ | 1.074x10$^{-2}$ | 2.909x10$^{-2}$ | 5.449x10$^{-2}$ | 1.264x10$^{-3}$ | α |
| $^{294}117$ | 11.124 | 7.705x10$^{7}$ | 1.538x10$^{3}$ | 6.773x10$^{-2}$ | 1.020x10$^{-1}$ | 1.903x10$^{-1}$ | 3.894x10$^{-3}$ | α |
| $^{296}117$ | 10.744 | 8.454x10$^{5}$ | 6.013x10$^{1}$ | 7.047x10$^{-1}$ | 1.040x10$^{0}$ | 1.805x10$^{0}$ | 3.156x10$^{-2}$ | α |
| $^{298}117$ | 10.794 | 1.806x10$^{3}$ | 9.028x10$^{-1}$ | 4.743x10$^{-1}$ | 6.944x10$^{-1}$ | 1.334x10$^{0}$ | 2.188x10$^{-2}$ | α |
| $^{300}117$ | 10.694 | 7.473x10$^{-1}$ | 5.364x10$^{-3}$ | 8.491x10$^{-1}$ | 1.227x10$^{0}$ | 2.449x10$^{0}$ | 3.656x10$^{-2}$ | SF |
| $^{302}117$ | 11.424 | 5.985x10$^{-5}$ | 1.306x10$^{-5}$ | 8.278x10$^{-3}$ | 1.209x10$^{-2}$ | 3.489x10$^{-2}$ | 5.732x10$^{-4}$ | SF |
| $^{304}117$ | 11.764 | 9.282x10$^{-10}$ | 1.349x10$^{-8}$ | 1.068x10$^{-3}$ | 1.559x10$^{-3}$ | 5.522x10$^{-3}$ | 9.226x10$^{-5}$ | SF |
| $^{306}117$ | 10.944 | 2.788x10$^{-15}$ | 6.120x10$^{-12}$ | 1.387x10$^{-1}$ | 1.943x10$^{-1}$ | 5.444x10$^{-1}$ | 6.905x10$^{-3}$ | SF |
| $^{308}117$ | 10.464 | 1.624x10$^{-21}$ | 1.261x10$^{-15}$ | 3.022x10$^{0}$ | 4.126x10$^{0}$ | 1.024x10$^{1}$ | 1.090x10$^{-1}$ | SF |
| $^{310}117$ | 9.244 | 1.833x10$^{-28}$ | 1.220x10$^{-19}$ | 2.405x10$^{4}$ | 3.211x10$^{4}$ | 4.842x10$^{4}$ | 3.937x10$^{2}$ | SF |

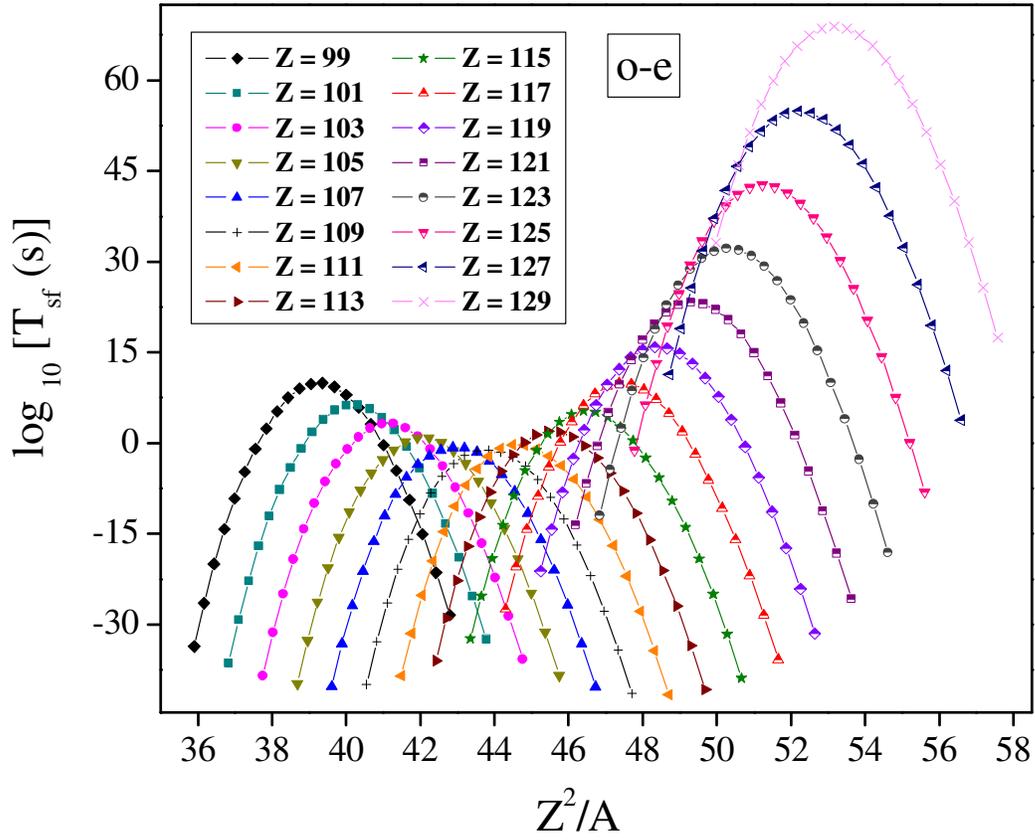

**Figure 9.** (Color online) Plot connecting the logarithmic spontaneous fission half lives against the fissionability parameter ($Z^2/A$) of odd-even isotopes with Z = 99 - 129.

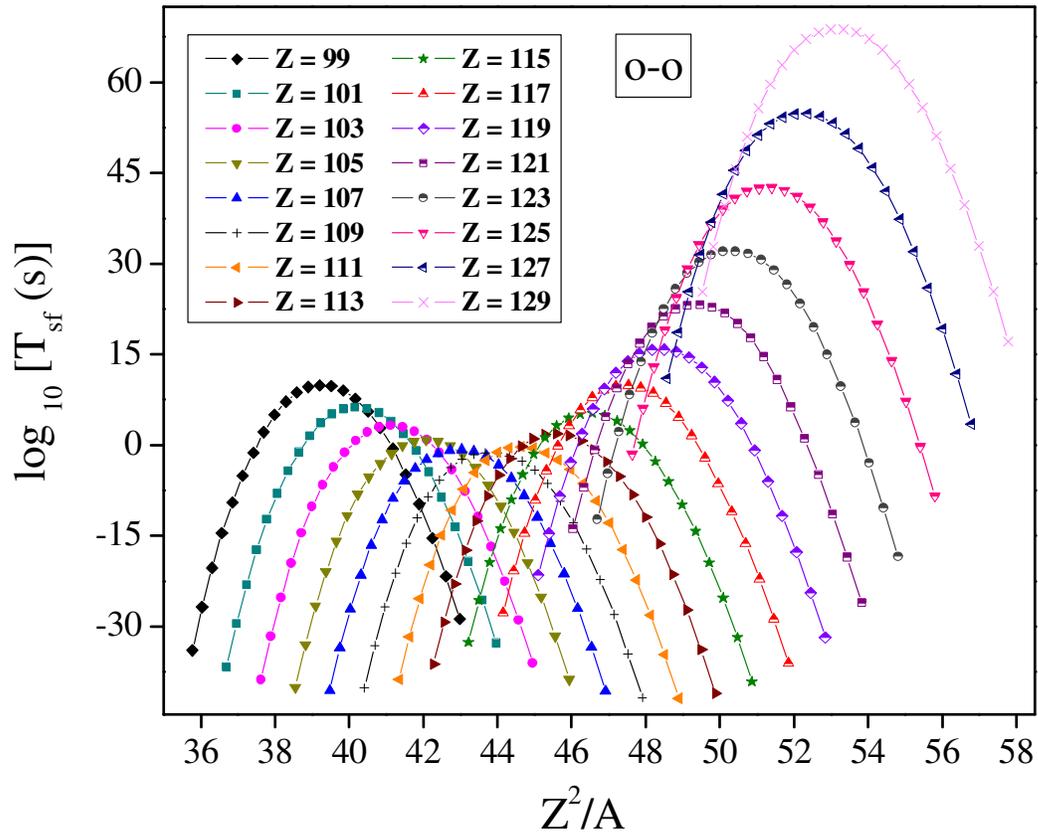

**Figure 10.** (Color online) Plot connecting the logarithmic spontaneous fission half lives against the fissionability parameter ($Z^2/A$) of odd-odd isotopes with Z = 99 - 129.

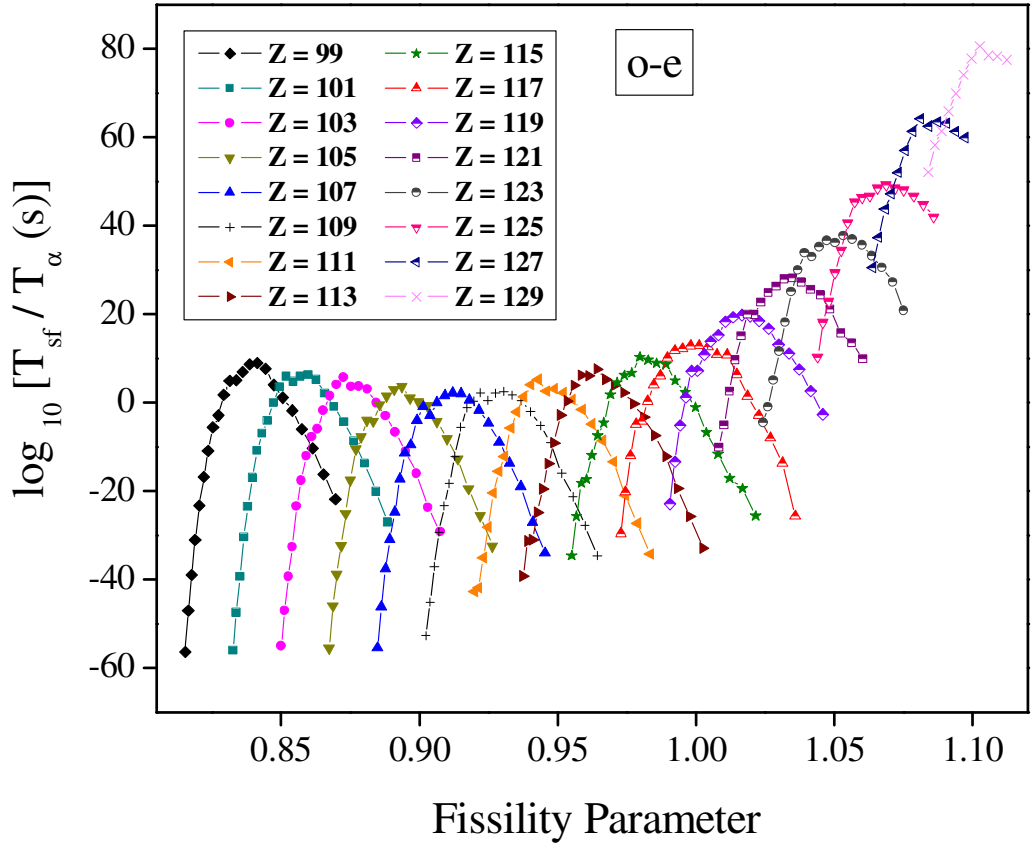

**Figure 11.** (Color online) Plot connecting the logarithmic ratio of the spontaneous fission half lives to the alpha half lives (CPPMDN) against the fissility parameter of odd-even isotopes with Z = 99 - 129.

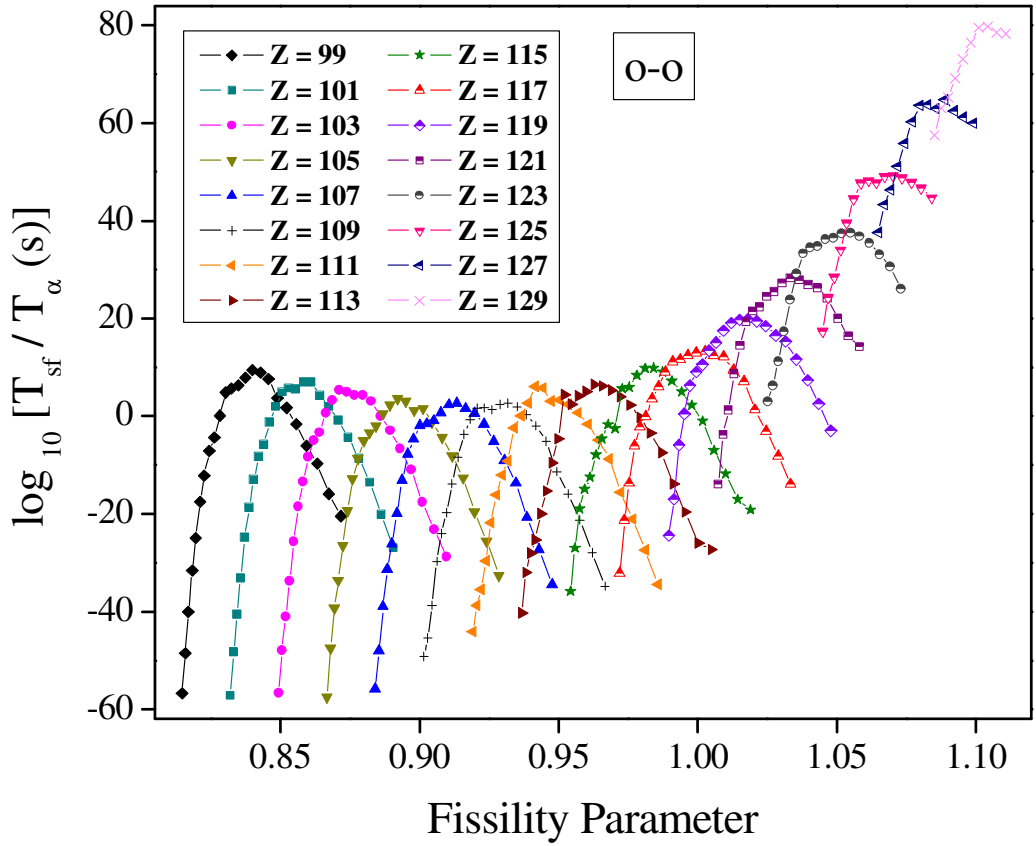

**Figure 12.** (Color online) Plot connecting the logarithmic ratio of the spontaneous fission half lives to the alpha half lives (CPPMDN) against the fissility parameter of odd-odd isotopes with Z = 99 - 129.

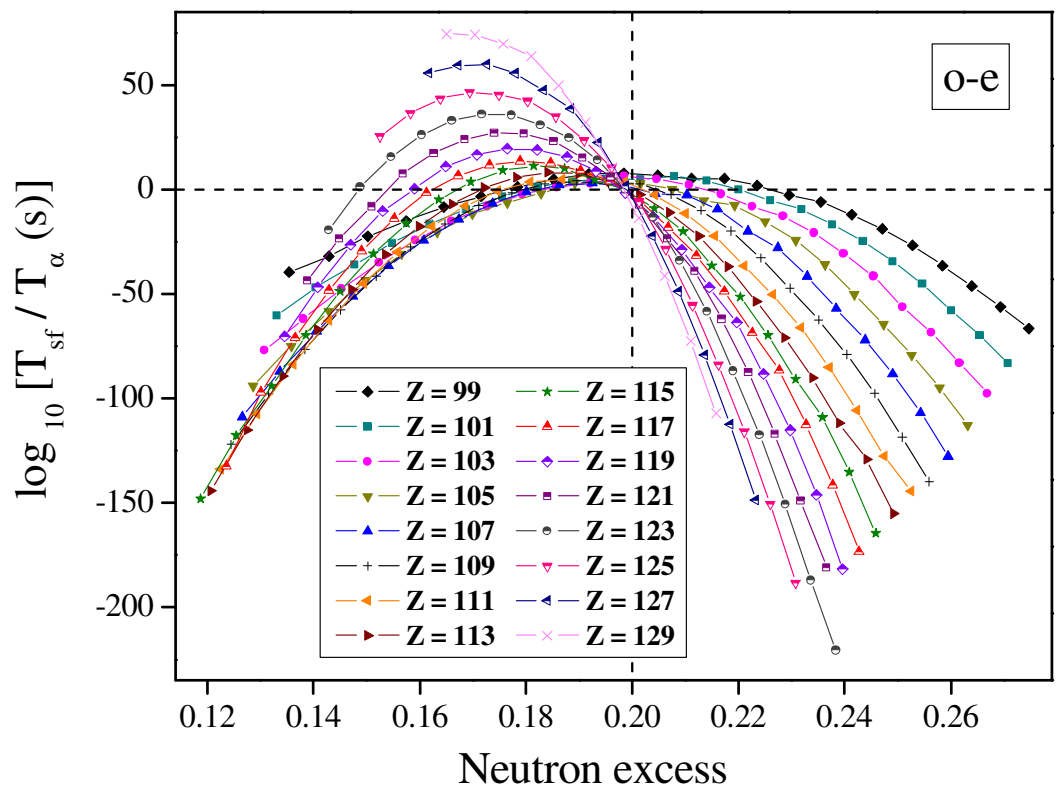

**Figure 13.** (Color online) Plot connecting the logarithmic ratio of the spontaneous fission half lives to the alpha half lives (CPPM) against the neutron excess of odd-even isotopes with Z = 99 - 129.

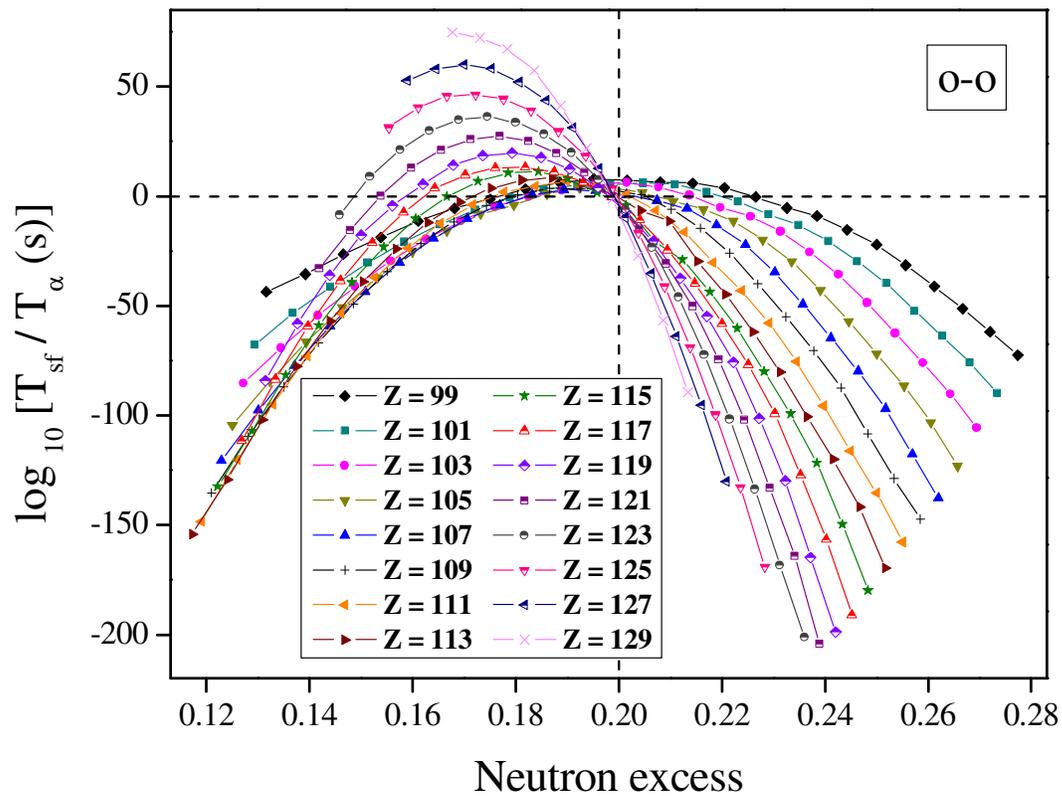

**Figure 14.** (Color online) Plot connecting the logarithmic ratio of the spontaneous fission half lives to the alpha half lives (CPPM) against the neutron excess of odd-odd isotopes with Z = 99 - 129.